\documentclass[english,reprint,amsmath,amssymb,aps,superscriptaddress,floatfix]{revtex4-1}
\usepackage[T1]{fontenc}
\usepackage[latin9]{inputenc}
\usepackage{color}
\usepackage{babel}
\usepackage{float}
\usepackage{textcomp}
\usepackage{amsmath}
\usepackage{amssymb}
\usepackage{graphicx}
\usepackage{wasysym}
\usepackage{array}
\usepackage[unicode=true,pdfusetitle,
 bookmarks=true,bookmarksnumbered=false,bookmarksopen=false,
 breaklinks=false,pdfborder={0 0 0},pdfborderstyle={},backref=false,colorlinks=true]
 {hyperref}
 
\makeatletter

\newcommand{\thickhline}{%
	\noalign {\ifnum 0=`}\fi \hrule height 1pt
	\futurelet \reserved@a \@xhline
}
\newcolumntype{"}{@{\hskip\tabcolsep\vrule width 1pt\hskip\tabcolsep}}

\newcommand*\LyXZeroWidthSpace{\hspace{0pt}}
\newcommand{\lyxmathsym}[1]{\ifmmode\begingroup\def\b@ld{bold}
  \text{\ifx\math@version\b@ld\bfseries\fi#1}\endgroup\else#1\fi}

\usepackage[caption=false]{subfig}
\usepackage{multirow}

\@ifundefined{showcaptionsetup}{}{%
 \PassOptionsToPackage{caption=false}{subfig}}
\usepackage{subfig}
\makeatother

\newlength{\uppercaseHeight}
\newcommand{\SC}{\text{SC}}

\begin{document}
\title{Charge order and Mott insulating ground states \\ in small-angle
twisted bilayer graphene}
\author{Markus J. Klug}
\email{markus.klug@kit.edu}

\affiliation{Institut für Theorie der Kondensierten Materie, Karlsruher Institut
für Technologie, 76131 Karlsruhe, Germany}
\begin{abstract}
In this work, we determine states of electronic order of small-angle
twisted bilayer graphene. Ground states are determined for weak and
strong couplings which are representatives for varying distances of the
twist-angle from its magic value. In the weak-coupling regime, charge
density waves emerge which break translational and $C_{3}$-rotational
symmetry. In the strong coupling-regime, we find rotational and translational
symmetry breaking Mott insulating states for all commensurate moir\'e
band fillings. Depending on the local occupation of superlattice sites
hosting up to four electrons, global spin-(ferromagnetic) and valley
symmetries are also broken which may give rise to a reduced Landau level
degeneracy as observed in experiments for commensurate
band fillings. 
The formation of those particular
electron orders is traced back to the important
role of characteristic non-local interactions which connect all localized states
belonging to one hexagon formed by the AB- and BA-stacked regions of the superlattice. 
\end{abstract}
\maketitle
The temperature -- gate voltage phase diagram of twisted bilayer graphene
(TBG) in the small-angle regime is characterized by correlated insulator
states as seen in transport experiments \citep{Cao18:PabloExpInsulatorAtHalfFill,Cao18:PabloExpSuperconductivity,Yankowitz19:DeanExpPressure,Lu10:EfetovExp,Polshyn19:ExpLinearRes}.
Their regular pattern of occurrences at commensurate fillings of the
weakly dispersing moir\'e bands with bandwidths as small as $10\text{meV}$
\citep{Bistritzer07:TBGContinuumModel} indicates an enhanced role
of interaction effects, including strong-coupling Mott physics complemented
by other complex electron phenomena such as superconductivity \citep{Cao18:PabloExpSuperconductivity,Cao18:PabloExpInsulatorAtHalfFill,Yankowitz19:DeanExpPressure,Lu10:EfetovExp,Polshyn19:ExpLinearRes},
linear-in-temperature resistivity \citep{Polshyn19:ExpLinearRes},
correlated electron states observed in scanning tunneling microscopy (STM) and scanning tunneling spectroscopy (STS) measurements \citep{Kerelsky18:PasupathyExp,Choi19:STM,Jiang19:STMChargeOrder,Xie19:ExpSTM},
ferromagnetism and quantum hall physics \citep{Sharpe19:ExpFerroAtThreeQuarter}.
Though vast theoretical efforts were made to model the electronic
structure \citep{Yuan18:LiangFusFirstTwoOrbitalModel,Koshino18:LiangMaxLocWO,Kang18:LocWannierStates,Po18:SenthilTheoryPaper,Zou18:Senthil2ndTheory,Carr19:TenBandModel,Po19:TightBindingModels,Goodwin19:InteractionAsFunctionofTwistAngle},
as well as to explain the superconducting pairing mechanism \citep{Laksono18:SCchargeOrder,Liu18:ChiralSDW,Guo18:PairingSym,Xu18:TopSC,Thomson18:AFMmitMathias,Wu18:PhononSC,Huang18:MonteCarloSCMott,Peltonen18:SCmeanfield,Kennes18:StrongCorrSC,Kozii19:NematicSCDensityFluct,Roy19:SCinTBG,Gonzales19:SCKohnLuttinger,You19:SCValleyFluct,Scheurer19,Lian19:PhononSC}
and the insulating states \citep{Isobe18:RG,Dorado18:CorInsulatorStateKivelson,Ochi2018:InsulatorState,Po18:SenthilTheoryPaper,Xu18:QMCKekule,Venderbos19:CorrRafael,Kang19:StrongCouplingPhase,Seo19:FerroMott,Rademaker19:ChargeTransfer},
a comprehensive understanding of the insulating states for variable
carrier concentrations is lacking. This is obviously important if
one wants to identify the mechanism of superconductivity in these
systems. 

\begin{figure}[t]
\centering{}\includegraphics[draft=false,width=1\columnwidth]{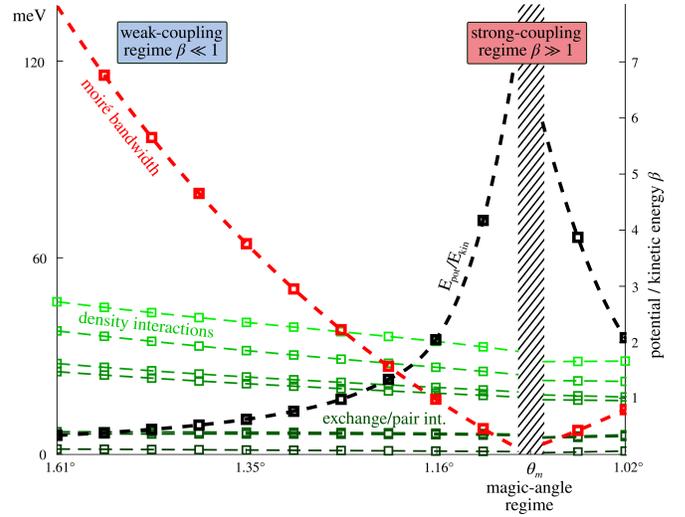}\caption{\label{fig:tbg-energetics}By tuning the twist-angle, twisted bilayer
graphene undergoes a transition from a weak- into a strong-coupling
regime as indicated by the ratio of potential to kinetic energy scale $ \beta $. 
The moir\'e bandwidth (red) represents the kinetic energy
scale whereas the amplitudes of the computed interaction matrix elements (green), which are determined in Sec.~\ref{sec:microscopic-model}, the potential energy scale. Their ratio (black) is strongly enhanced when tuning the system towards
the magic-angle regime indicating a crossover from weak to strong couplings. }
\end{figure}

Quantum oscillations reveal that the insulating states differ in their
Landau level degeneracy \citep{Cao18:PabloExpSuperconductivity,Cao18:PabloExpInsulatorAtHalfFill,Yankowitz19:DeanExpPressure,Lu10:EfetovExp}.
This indicates the presence (or absence) of (global) symmetries which
generate Kramer-like degeneracies of single-particle states. 
By assuming 
the presence of a spin-rotation and a valley symmetry, 
which is justified in the limit of vanishing spin-orbit coupling \citep{Neto09:ReviewGraphene}
and small twist-angles \citep{Po18:SenthilTheoryPaper}, 
this observation may be interpreted as follows: 
It implies for the insulator state at charge neutrality ($\nu=0$) which is
found to be $4$-fold degenerate the presence of both the spin and valley symmetry,
for half electron- or hole-filling ($\nu=\pm1/2$) which are two-fold
degenerate that either the spin or the valley symmetry is broken, and for the
band fillings $\nu=\pm3/4$ which are single degenerate that both the spin
and the valley symmetry are absent \citep{Cao18:PabloExpSuperconductivity,Yankowitz19:DeanExpPressure,Lu10:EfetovExp}. 
Here, the band filling factors $ \nu = -1 ,1 $ represent completely empty and completely filled moir\'e bands, respectively.

In addition to transport experiments \citep{Cao18:PabloExpSuperconductivity,Cao18:PabloExpInsulatorAtHalfFill,Yankowitz19:DeanExpPressure,Lu10:EfetovExp,Polshyn19:ExpLinearRes},
which are performed on samples identified with the so-called \textit{magic-angle
regime} \citep{Bistritzer07:TBGContinuumModel} hosting among others
correlated insulator states and superconductivity, STS and STM experiments
reveal correlated electron states for moir\'e band fillings around the 
charge neutrality point (CNP) \citep{Kerelsky18:PasupathyExp,Choi19:STM,Jiang19:STMChargeOrder,Xie19:ExpSTM}.
Here, interaction effects manifest themselves as a significant redistribution
of the single-particle spectral weight which sets in at a critical amount of electron-
or hole-doping. 
Additionally, it is unanimously reported that the correlated states break $C_{3}$-rotational 
symmetry as seen in spatially resolved charge distribution measurements  \citep{Kerelsky18:PasupathyExp,Choi19:STM,Jiang19:STMChargeOrder,Xie19:ExpSTM},
and that this effect is largest at the CNP. These results are obtained
for samples with moir\'e bandwidths significantly
larger than 10meV which may indicate a placement of these samples 
in a \textit{close-to-magic-angle regime} 
where correlation effects are present yet with relative interaction strengths significantly smaller than those representative for 
the magic-angle regime. 

This may be due to the fact that the 
magic-angle regime is in general not determined by the twist-angle $ \theta $ but also by the interlayer coupling $ t_\bot $. 
The condition for the magic-angle regime can be formulated in terms of the dimensionless quantity $\alpha= \frac{t_{\bot}}{\hbar v_{F} k_\theta}$ with the Fermi velocity $v_{F}$ and the relevant inverse length scale given by the distance between the Dirac points of the twisted graphene systems in the reciprocal space $ k_\theta = \frac{8\pi }{3 a} \sin (\theta/2) $ with the graphene lattice constant $ a $.  
In the magic-angle regime, the dimensionless quantity takes on the value $\alpha_{m}\approx 1/ \sqrt{3} $ \citep{Bistritzer07:TBGContinuumModel}. 
Hence, the role of interaction effects, which are assumed to be enhanced for small bandwidths, 
depends on $\theta$ and $t_{\bot}$ which may be different for different types of experimental settings. 

In the subsequent analysis, we find that the relative strength of interaction effects in our model for TBG is determined by the 
dimensionless ratio of interaction and kinetic energy scales which is given by 
\begin{equation}
\label{eq:beta}
\beta \equiv \frac{e^2}{\epsilon L_M \Lambda} 
\end{equation}
with the electron charge $ e $ and the relative permittivity $ \epsilon $ determined by the substrate. 
Here, the moir\'e bandwidth $ \Lambda = \Lambda (\theta, t_\bot) $, which is minimal in the magic-angle regime,  depends on the twist-angle and the interlayer coupling, whereas the potential energy scale $ e^2 / \epsilon L_M  $ represents the strength of electron-electron interactions and scales with the inverse of the characteristic length of the superlattice $ L_M = \frac{4\pi}{3} k_\theta^{-1} $. 
As $ \Lambda $ and $ L_M $ scale differently with the twist-angle, the system is expected to crossover from a strong-coupling regime with $ \beta \gg 1 $ identified with the magic-angle regime to a weak-coupling regime with $ \beta \ll 1 $ identified with the close-to-magic-angle regime by tuning the twist-angle (or equivalently the interlayer coupling). 
An illustration of this expected crossover is presented in Fig.~\ref{fig:tbg-energetics}. 

\begin{figure}[b]
\centering{}\includegraphics[draft=false,width=1\columnwidth]{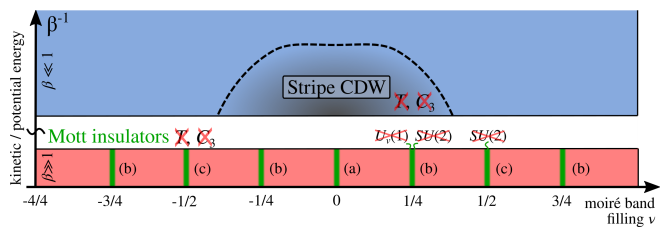}\caption{\label{fig:results}In the \textit{weak-coupling regime} ($ \beta \ll 1 $, bluish shaded background), our ground
state analysis reveals a formation of a stripe charge density wave order
which breaks the translational $T$ and the $C_{3}$-rotational symmetry of the lattice. 
The line of the second order phase transition is represented by the dashed line. 
In the \textit{strong-coupling regime} ($ \beta \gg 1 $, reddish shaded background), we find three types of Mott
insulating ground states (green lines a, b, c) for all commensurate moir\'e band
fillings $\nu=0,\pm1/4,\pm1/2,\pm3/4$, which resemble the stripe-type
orders in the weak-coupling limit (a) but differ, in part, by the
absence of the spin $SU(2)$ (c) as well as the valley $U_{v}(1)$ symmetry
(b). }
\end{figure}

In this work, we investigate possible electronic
ground states in the presence of Coulomb electron-electron
interaction as a function of the moir\'e band filling and the twist-angle. We will distinguish between two
parameter regimes, a strong-coupling regime, $ \beta \gg 1$, representative for angles in the vicinity of the magic-angle
where interaction effects dominate, and a weak-coupling
regime, $ \beta \ll 1$, where interaction effects are subordinate.
As outlined above, both regimes are likely to be realized in experiments and deserve thorough investigation.

To this end we set up a single-particle continuum theory for
TBG as discussed in Refs.~\citep{Bistritzer07:TBGContinuumModel,Weckbecker16:TBGContinuumModel}
with superlattice translational $T$, crystalline point group $D_{6}$,
spin $SU(2)$, and valley conservation $U_{v}(1)$ approximate symmetries
\citep{Zou18:Senthil2ndTheory}. Interaction effects are considered
by employing a two-orbital model introduced in Refs.~\citep{Yuan18:LiangFusFirstTwoOrbitalModel,Koshino18:LiangMaxLocWO}
where maximally localized Wannier functions centered at the AB- and BA-stacked
regions of the superlattice are constructed from the moir\'e
bands. Subsequently, interaction matrix elements between the Wannier
states in the direct, exchange and pair-hopping channel are computed
which are found to be highly non-local. 
By identifying the relevant interaction processes, 
a minimal tight binding model of interacting moir\'e electrons is obtained eventually which forms the basis of the following analysis. 

In the \textit{weak-coupling regime} ($\beta \ll 1$), our results, which are obtained
in an unrestricted mean field analysis, reveal a formation of a stripe
charge density wave order with commensurate ordering vectors of half
of the reciprocal lattice vectors of the moir\'e superlattice which break translational $ T $ and $ C_3 $-rotational symmetry. The
charge inhomogeneities form when the interaction strength and the
moir\'e band filling reach a certain critical threshold which is determined by the usual criterion for the onset of long-range orders in mean field theories. 
The line of transition as function of relative interaction strength and moir\'e band filling is depicted in ~Fig.~\ref{fig:results}. 
In fact, we find no weak coupling instability which could be due to the presence of a nesting condition 
or which could be connected to the diverging density of states at the van Hove points of the single-particle spectrum. 

In the \textit{strong-coupling regime} ($\beta \gg 1$), we perform an "infinite coupling"
limit where we drop the kinetic part of the theory and consider only
interaction processes which can be expressed in terms of density-density
interactions. This approach would be futile in the
limit of local Hubbard interactions. However, for the present model,
the important role of non-locality combined with the ferromagnetic and
ferrovalley exchange interactions distinguishing between spin and
valley numbers allows us to determine the nature of the ordered states
even without the kinetic energy contributions of the electrons. For all commensurate
band fillings $\nu=0,\pm1/4,\pm1/2,\pm3/4$, we find Mott insulating
ground states which break the translational $T$ and $C_{3}$-rotational 
symmetry ($\nu=0$), as well as the spin $SU(2)$  ($\nu=\pm1/4,\pm1/2,\pm3/4)$
and valley $U_{v}(1)$ symmetry ($\nu=\pm1/4,\pm3/4$). 
The results are summarized in Fig.~\ref{fig:results}. 

\section{Microscopic model}
\label{sec:microscopic-model}

We first obtain the weakly
dispersing moir\'e bands by considering a continuum model \citep{Bistritzer07:TBGContinuumModel,Weckbecker16:TBGContinuumModel}
where states near the slightly twisted Dirac cones of the two graphene
layers hybridize due to a finite inter-layer coupling. 
Because of
the large separation in momentum space, states near non-equivalent
cones, in the following labelled by the valley quantum number $\xi=\pm$,
are assumed to be effectively decoupled generating an emergent $U_{v}(1)$
valley symmetry. 

The single-particle Hamiltonian for TBG expressed
in the two-layer graphene basis, $\phi = (\phi_A^{(1)},\phi_B^{(1)},\phi_A^{(2)},\phi_B^{(2)})$ with the crystalline sublattice basis labelled by the indices $ A $ and $ B $, is written as
\citep{Weckbecker16:TBGContinuumModel} 
\begin{equation}
H_{TBG}=\sum_{\mathbf{K}\mathbf{K}'\sigma\mathbf{\xi}}\phi_{\mathbf{K}\xi\sigma}^{\dagger}\begin{pmatrix}H_{\xi,\theta/2}\!-\!\mu & T(t_\bot) \\
T^{\dagger} (t_\bot) & H_{\xi,-\theta/2}\!-\!\mu
\end{pmatrix}_{\mathbf{K}\mathbf{K'}}\phi_{\mathbf{K}'\xi\sigma},\label{eq:Htbg}
\end{equation}
where $H_{\xi,\varphi} $ denotes the Hamiltonian describing single electrons in the single
graphene layer near valley $\xi$ which is rotated by the angle $\varphi$ and the chemical potential $ \mu $.
The interlayer coupling is described by the matrix $T (t_\bot)$ which depends on the interlayer tunneling amplitude $ t_\bot $. As parameters, we choose
the Fermi velocity to $v_{F}\hbar/a=2.1354$eV and the inter-layer
tunneling amplitudes, which discriminate between intra- and intersublattice
processes, to $t_{\bot,AA}=t_{\bot,BB}=79.7$meV and $t_{\bot,AB}=97.5$meV, to take
lattice relaxation effects into account following Ref.~\citep{Koshino18:LiangMaxLocWO}.
$\mathbf{K}$ denotes the crystal momentum in the single layer graphene
Brillouin zone and $\sigma$ the electron spin. 
Subsequently, by diagonalizing Eq.~\eqref{eq:Htbg} and performing a back folding of the electronic states into the moir\'e Brillouin zone, the
effective Hamiltonian describing the flat moir\'e bands is obtained by considering only the
four narrow bands around charge neutrality. It is given by 
\begin{equation}
H_{M}=\sum_{\lambda\mathbf{k}\sigma\xi}\left(\epsilon_{\lambda\mathbf{k}\xi}\!-\!\mu\right)\psi_{\lambda\mathbf{k}\xi\sigma}^{\dagger}\psi_{\lambda\mathbf{k}\xi\sigma} \label{eq:Hm}
\end{equation}
with the dispersion relation $\epsilon_{\lambda\mathbf{k}\xi}$, and where the band index $\lambda\in\{1,2\}$, the spin index $\sigma$, the valley index $\xi$ and the crystal
momentum $\mathbf{k}$, which is element of the moir\'e Brillouin zone, label
the superlattice Bloch states $\psi_{\lambda\mathbf{k}\sigma\xi}$. 
The bandwidth of the moir\'e bands, $ \Lambda \equiv  \max \epsilon_{\lambda = 1} - \min \epsilon_{\lambda=2} $ where $ \lambda=1 $ ($ \lambda = 2 $) labels the conduction (valence) band, is depicted as function of the twist-angle in Fig.~\ref{fig:tbg-energetics} with a magic-angle determined to $ \theta_m \approx 1.08\lyxmathsym{\protect\textdegree} $, 
whereas a representation of the moir\'e band structure for a particular twist-angle is found in the Appendix \ref{subsec:hopping-parameter}.

\subsection{Construction of the Wannier basis}
The Wannier basis is constructed by employing a two-orbital model
\citep{Po18:SenthilTheoryPaper,Kang18:LocWannierStates,Koshino18:LiangMaxLocWO}
where the localized Wannier functions, though centered at the AB- and
BA-stacked regions of the superlattice, possess highest weight at the AA-stacked regions. Despite the occurrence of a Wannier
obstruction, which renders certain exact symmetries non-local \citep{Po18:SenthilTheoryPaper,Zou18:Senthil2ndTheory}
and is resolved by incorporating auxiliary bands \citep{Carr19:TenBandModel,Po19:TightBindingModels}
which add a vast number of degrees of freedom, we assume that the relevant
physics of our work is captured by the two-orbital model. 
In particular, we follow the approach outlined in Ref.~\citep{Koshino18:LiangMaxLocWO}
and construct Wannier functions centered at the AB- and BA-stacked regions by employing the method of maximally
localized Wannier functions \citep{Marzari97:MaxLocWanFunc,Marzari12:ReviewMaxLocWannierFunctions}.
The Wannier states possess a definite valley number whereas the associated Wannier functions of different valleys are connected by complex conjugation because of the presence of an effective time reversal symmetry \cite{Koshino18:LiangMaxLocWO,Po18:SenthilTheoryPaper}.
Details about our construction of the Wannier basis are found in the Appendix \ref{sec:Construction-of-wannier}. 

Eventually, we obtain orthogonal
and exponentially localized Wannier states $\Psi_{i\alpha\xi\sigma}\left(\mathbf{r}\right)$ 
located in the superlattice unit cell $i$ and centered
at the $\alpha=\text{AB,BA}$-stacked regions of the superlattice as well as labelled by the valley $ \xi $ and spin $ \sigma $ numbers. 
This set of states constitutes the single-particle Wannier basis which gives rise to
8 states per superlattice unit cell.
By considering generic two-particle interactions between the Wannier states, the effective tight-binding model
describing interacting moir\'e electrons is given by
\begin{multline}
H=\sum_{ab\sigma}(t_{a,b}-\mu\delta_{ab})c_{a\sigma}^{\dagger}c_{b\sigma} \\
+\tfrac{1}{2}\sum_{\sigma\sigma'}\sum_{abcd}U_{abcd}c_{a\sigma}^{\dagger}c_{b\sigma'}^{\dagger}c_{c\sigma'}c_{d\sigma},\label{eq:eff-model}
\end{multline}
where $ c_{a\sigma}^{(\dagger)} $ annihilates (creates) an electron with spin $ \sigma $ in the Wannier state $ a=(i,\alpha,\xi) $. The transition amplitudes $t_{i\alpha\xi,i'\alpha'\xi'} = \delta_{\xi\xi'} t_{i\alpha,i'\alpha'} $, which are by construction diagonal
in the valley indices, are obtained by expressing the Hamiltonian Eq.~\eqref{eq:Hm} in the constructed Wannier basis and reproduce the flat moir\'e bands in the reciprocal space as shown in the Appendix \ref{subsec:hopping-parameter}.  
The interaction matrix elements $ U_{abcd} $ are determined in the subsequent section. 

\begin{figure}[b]
\centering{}\includegraphics[width=1\columnwidth]{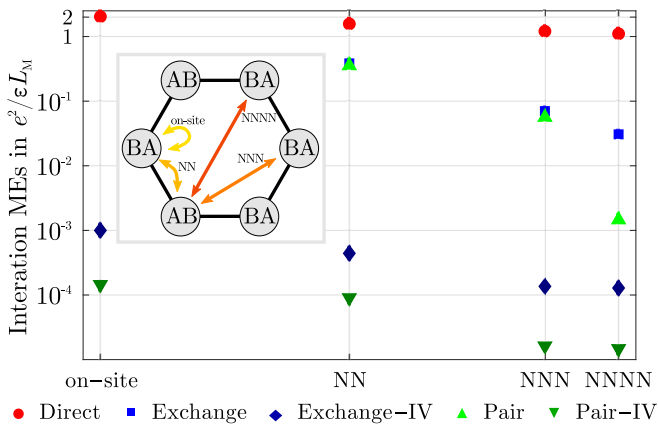}\caption{\label{fig:interaction-ME}Amplitudes of interaction matrix elements
in units of $e^{2}/\epsilon L_{\text{M}}$ obtained for $\theta=1.16\lyxmathsym{\protect\textdegree}$ by evaluating Eq.~\eqref{eq:ME}.
All interaction processes which connect superlattice sites belonging to one hexagon formed by the AB- and BA-stacked regions of the superlattice
are found to be relevant. Longer-distance interactions are numerically small
and therefore negligible. Since the twist-angle dependence of
the interaction matrix elements is rather weak and approximately determined
by the superlattice length scale $U_{abcd}\propto L_{\text{M}}^{-1} $, the
numerical values are representative for both the strong- and the weak-coupling
regime. }
\end{figure}

\subsection{Interaction matrix elements}
\label{subsec:interaction-ME}
We compute the interaction matrix elements,
which are part of the effective Hamiltonian Eq.~(\ref{eq:eff-model}) describing two-particle interactions 
between Wannier states, by using an unscreened Coulomb kernel,
\begin{equation}
U_{abcd}=\frac{e^{2}}{4\pi\epsilon}\int_{{\bf r}{\bf r}'}\frac{\Psi_{a\sigma}^{\dagger}\left(\mathbf{r}\right)\Psi_{b\sigma'}^{\dagger}\left(\mathbf{r}'\right)\Psi_{c\sigma'}\left(\mathbf{r}'\right)\Psi_{d\sigma}\left(\mathbf{r}\right)}{|\mathbf{r}-\mathbf{r}'|},\label{eq:ME}
\end{equation}
with electron charge $e$ and relative permittivity $\epsilon\approx7$
for hexagonal boron nitride (hBN). We distinguish between density ($a=d$
and $b=c$ introducing $U=U_{abba}$), exchange ($a=c\neq b=d$ introducing
$J=U_{abab}$) and pair-hopping ($a=b\neq c=d$ introducing $X=U_{aabb}$) interaction processes.
We also examined charge-bond interaction matrix elements ($a\neq d\neq b=c$),
but find them at least one order of magnitude smaller than the previously introduced 
matrix elements. These processes are therefore safely neglected. 
Generally, we distinguish
between intravalley processes, where the valley indices in $a=(i,\alpha,\xi)$ and
$b=(i',\alpha',\xi)$ are identical, and intervalley processes, where the valley indices
in $a=(i,\alpha,\xi)$ and $b=(i',\alpha',-\xi)$ differ. The latter are labelled by the subscript $IV$
in the following. 
Note that the interaction matrix elements in the density channel depend on the absolute square of the single-particle wave functions rendering the distinction between intra- and intervalley processes obsolete as the wave functions of the different valleys are connected by complex conjugation. 

The results for the various interaction matrix elements which are obtained in a numerical evaluation of Eq.~\eqref{eq:ME} are depicted for one particular twist-angle in Fig.~\ref{fig:interaction-ME}. 
We find that the amplitude of interaction matrix elements drops with
distance between interacting Wannier states but remains significant for processes connecting all states
which belong to one hexagon formed by the AB- and BA-stacked regions of the
superlattice as depicted in the inset. 
This observation is traced back to the fact that the shape of the Wannier functions is highly non-local
with substantial overlap of Wannier functions of neighboring sites. 
In contrast, longer-distance interactions processes are numerically smaller due to the absence of a direct overlap of Wannier functions. 
This trend is further enhanced by screening effects which are present due to the short distance of the TBG sample to the back gate which can be of order of the superlattice unit cell. 
These interaction processes are therefore neglected in the subsequent analysis.  

Furthermore, we find that the dependence of the matrix elements on the twist-angle is rather weak 
and determined to leading order by the superlattice  constant $ U_{abcd} \propto L_M^{-1} \propto  \sin( \theta/2) $ as depicted in Fig.~\ref{fig:tbg-energetics} and discussed in detail in the Appendix \ref{subsec:interaction-matrix-elements}.
Apparently, as $ L_M $ is the characteristic length scale of the Wannier functions, it also sets the characteristic length scale of the interaction processes
justifying the usage of $ \beta $ defined in Eq.~\eqref{eq:beta} for a range of twist-angles around the magic-angle regime to characterize interaction effects. 
Additionally, the effect of screening is investigated in more detail. 
As the distance of the TBG sample to the metallic back gate is determined by the thickness of the hBN layer, which ranges between $ 10\dots30\text{nm} $ \citep{Cao18:PabloExpSuperconductivity,Cao18:PabloExpInsulatorAtHalfFill}, screening effects are expected to be relevant for superlattice unit cell sizes of order of this  distance. E.g.~in the vicinity of the magic-angle regime, $ L_M (\theta \negmedspace = \negmedspace 1.12 \lyxmathsym{\protect\textdegree})\negmedspace\approx\negmedspace13\text{nm} $. 
To this end, the interaction potential in Eq.~\eqref{eq:ME} is replaced by a Yukawa-type potential parametrized by a screening length which is chosen to match the distance between TBG sample and metallic back gate as discussed in the Appendix \ref{subsec:interaction-matrix-elements}. 
We find that screening alters the twist-angle dependence of the interaction matrix elements 
where non-local interaction processes are more strongly suppressed for larger $ L_M $ than local interaction processes. 
These findings are in agreement with results presented in Ref.~\cite{Goodwin19:InteractionAsFunctionofTwistAngle} where the effect of screening was investigated for a different type of screening potential.
However, up to an overall change of the amplitudes, which can be compensated in a redefinition of $ \beta $, 
the quantitative changes of the ratio of the various interaction elements are small for twist-angles in the vicinity of the magic-angle and are found to not affect the results of the subsequent analysis.
This is traced back to the fact that the predominant contribution to the interaction matrix elements arises from the areas of a direct overlap of Wannier functions where screening is inefficient. 
Hence, the strength of interaction processes is reasonably well specified by the parameter $ \beta $ and the results for the interaction matrix elements which are depicted in Fig.~\ref{fig:interaction-ME} 
are representative for the weak- and the strong-coupling regime. 

In general by inspecting the numerical values of matrix elements depicted in Fig.~\ref{fig:interaction-ME}, interactions are dominated by the direct interaction processes. Since
this type of interaction is insensitive to local valley or spin configurations,
we expect that possible charge modulations are determined by $U$.
Though at least one order smaller in amplitude but being sensitive
to valley and spin number, intra- and intervalley exchange processes $J$ and
$J_{IV}$, as well as intra- and intervalley pair-hopping $X$ and
$X_{IV}$ processes are expected to be relevant. Due to rapid phase
fluctuations, intervalley processes are much smaller. However, because
of the coupling of otherwise decoupled valley sectors, they are considered
relevant to determine the exact ground state. For both the intra-
and intervalley case, all matrix elements are found positive $J,J_{IV}>0$
causing neighboring spin- and orbital-degrees of freedom to align.

Concluding, the large hierarchy of amplitudes of interaction matrix elements,
\begin{equation}
U\gg J>|X|\gg J_{IV}>|X_{IV}|,\label{eq:hierarchy-ME}
\end{equation}
is characteristic for small-angle TBG and will decisively determine the nature of the electronic ground states discussed in the next section. 

\section{Ground state analysis} 
\label{sec:ground-state-analysis}

The effective tight-binding model Eq.~(\ref{eq:eff-model}) with the relevant interaction processes identified in the previous section constitutes
the basis for the subsequent ground state analysis, which is two-fold:
We first investigate a weak-coupling regime where $ \beta \ll 1 $, which is representative for twist-angles of the close-to-magic angle regime,
and second conduct a strong coupling analysis where $ \beta \gg 1 $, which is representative for twist-angles in the vicinity of the magic-angle. 
As the dependence of the interaction matrix elements on the twist-angle is approximately determined by the characteristic length of the superlattice $ L_M $,  we consider a fixed ratio between the various interaction matrix elements given in Fig.~\ref{fig:interaction-ME}, which we consider representative for both regimes, and  
use the dimensionless quantity $ \beta $ to tune the effective strength of interactions. 

\subsection{Weak-coupling regime}
\label{sec:weak-coupling-regime}
In this section, the effect of interactions with weak coupling strengths, $ \beta \ll 1 $, is investigated. 
To this end, we conduct a mean field analysis to identify the electron interaction channel which first develops an instability. 
Here, it is not intended to identify the exact twist-angle at which a transition occurs as it depends on many microscopic parameters which are beyond the scope of this work. 
Instead, relevant for this discussion are the ratios between the various interaction matrix elements which were determined in the previous section and the dimensionless quantity $ \beta $ which is considered a small tuning parameter of the relative strength of interactions. 

To ensure that our analysis is susceptible to various kinds of
electron instabilities, the mean field decoupling of the interaction terms of Eq.~\eqref{eq:eff-model} is conducted locally in all possible channels. 
For a general interaction term, we obtain 
\begin{multline}
c_{a\sigma}^{\dagger}c_{b\sigma'}^{\dagger}c_{c\sigma'}c_{d\sigma}\approx\langle c_{b\sigma'}^{\dagger}c_{c\sigma'}\rangle c_{a\sigma}^{\dagger}c_{d\sigma}+\langle c_{a\sigma}^{\dagger}c_{d\sigma}\rangle c_{b\sigma'}^{\dagger}c_{c\sigma'}\\
-\langle c_{a\sigma}^{\dagger}c_{c\sigma'}\rangle c_{b\sigma'}^{\dagger}c_{d\sigma}-\langle c_{a\sigma}^{\dagger}c_{c\sigma'}\rangle c_{b\sigma'}^{\dagger}c_{d\sigma}+\text{const.}.\label{eq:mean fields}
\end{multline}
This decoupling scheme gives rise to a quadratic single-particle mean field Hamiltonian, 
\begin{align}
H&_{\text{MF}}  =  \sum_{ab\sigma\sigma'}\Big\{ \big( t_{ab}-\mu\delta_{ab} \notag \\
 &+ \sum_{cd\sigma''}\big[U_{adcb}\langle c_{d\sigma''}^{\dagger}c_{c\sigma''}\rangle+U_{cabd}\langle c_{c\sigma''}^{\dagger}c_{d\sigma''}\rangle\big]\big)\delta_{\sigma\sigma'} \notag \\ 
&-\sum_{cd}\big[U_{dacb}\langle c_{d\sigma'}^{\dagger}c_{c\sigma}\rangle+U_{dbca}\langle c_{d\sigma'}^{\dagger}c_{c\sigma}\rangle\big]\Big\} c_{a\sigma}^{\dagger}c_{b\sigma'}
,\label{eq:MF-ham}
\end{align}
with the local mean fields $ \langle c_{a\sigma}^{\dagger}c_{b\sigma'} \rangle  $ as variational parameters. Here, local correlations are straightforwardly determined by employing standard numerical methods such as the Lanczos algorithm \cite{Lanczos50}. 
The ground state is eventually determined in a self-consistency procedure by minimizing a ground state energy functional on a finite lattice of $30\times30$ superlattice unit cells. 
Details to the numerical computation scheme and a discussion of possible competing electronic orders are found in the Appendix \ref{sec:Lanczos-algorithm}. 

\begin{figure}[t]
	\centering{}\includegraphics[draft=false, width=1\columnwidth]{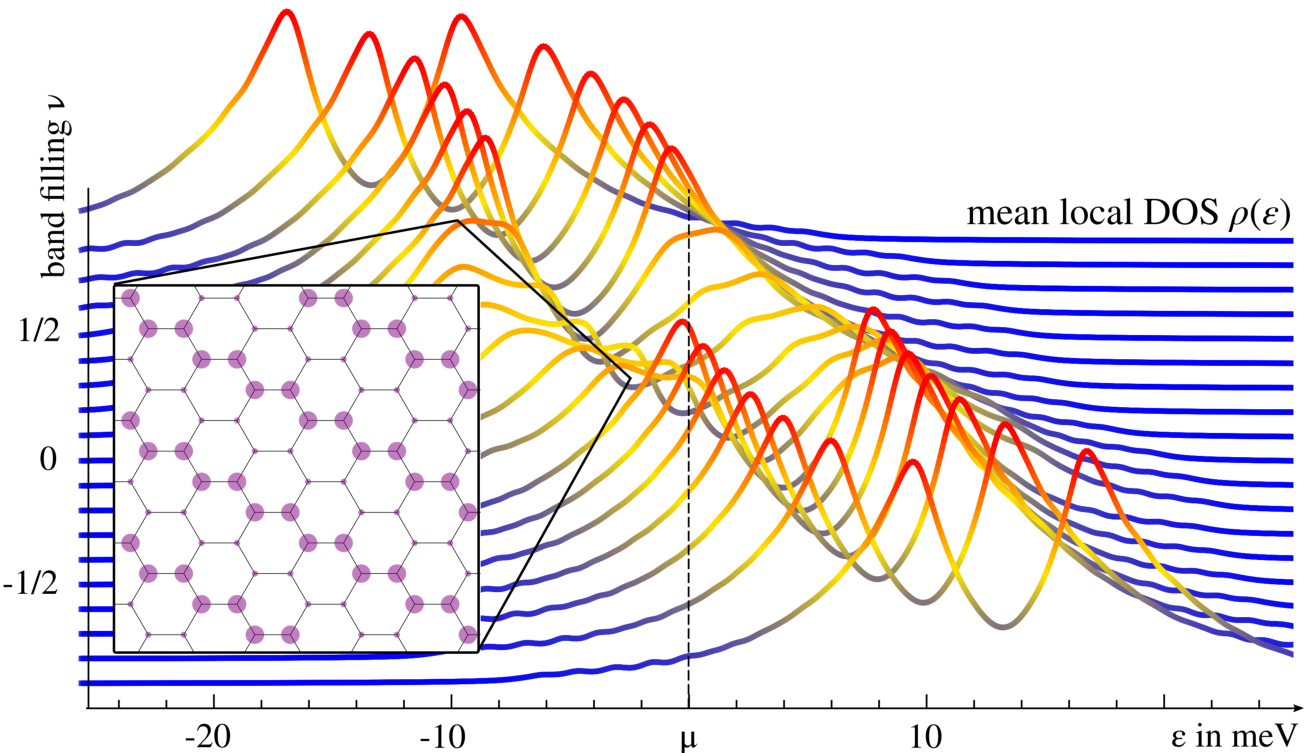}
	\caption{\label{fig:weak-coupling-results} The mean local density of states as
		function of frequency ($x$-axis) for various moir\'e band fillings
		($y$-axis) for a fixed twist-angle in the \textit{weak-coupling regime}. 
		The local density of states is averaged over all lattice sites and is given by $ \rho (\epsilon) = -\frac{1}{\pi N} \sum_{a,\sigma} \text{Im} G^R_{aa,\sigma\sigma}(\epsilon) $,
		where the retarded Green's function is obtained from the mean field Hamiltonian as given in Eq.~\eqref{eq:mean field-GF} and $ N $ denotes the number of lattice sites. 
		In the normal state, the single-particle spectrum is characterized by van Hove peaks as observable for high and low densities.
		The onset of a stripe density wave order causes a significant distortion of the single-particle
		spectrum, where the effect is largest around half filling. 
		A real space representation of the stripe charge density wave order with $ \Delta_\textbf{Q} / n \approx 0.126  $
		is depicted in the
		inset where the hexagon's vertices represent the AB- and BA-stacked
		regions of the superlattice. The purple dot's diameter scales with
		the local occupation of orbitals %
		\parbox[c][1\baselineskip]{0.55cm}{%
			\protect\includegraphics[height=0.8\baselineskip]{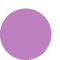}%
		}$\propto\sum_{\xi\sigma}\langle\hat{n}_{i\alpha\xi\sigma}\rangle$. }
\end{figure}

The interaction channel which first develops a long-range order at an effective interaction strength of $ \beta \gtrsim 0.04 $
is a stripe charge density wave which breaks translational $ T $ and
$C_{3}$-rotational symmetry while preserving the spin $ SU(2) $ and valley $ U(1) $ symmetry. 
Here, the local electron density is parametrized by 
\begin{equation}
\langle\hat{n}_{i\alpha\xi\sigma}\rangle = \tfrac{n}{8} + \tfrac{\Delta_\textbf{Q} }{8N}\cos(\mathbf{Q}\cdot\mathbf{R}_{i\alpha}),
\end{equation}
where the order parameter is given by
\begin{align}
\Delta_\textbf{Q} = \sum_{\textbf{k}\alpha\xi\sigma} \langle c^\dagger_{\textbf{k}+\textbf{Q}\alpha\xi\sigma  } c_{\textbf{k}\alpha\xi\sigma  } \rangle ,
\label{eq:orderPar}
\end{align}
with the electron density $ n \negmedspace = \negmedspace \tfrac{1}{N} \sum_{i\alpha\xi\sigma} \langle\hat{n}_{i\alpha\xi\sigma}\rangle $ and the number of superlattice unit cells $ N $, the particle number operator $\hat{n}_{i\alpha\xi\sigma} \negmedspace=\negmedspace c_{i\alpha\xi\sigma}^{\dagger}c_{i\alpha\xi\sigma}$
and the lattice site vector $\mathbf{R}_{i\alpha}$. Possible ordering vectors
are \textbf{$\mathbf{Q}\in\left\{ \mathbf{G}_{1}/2,\mathbf{G}_{2}/2,(\mathbf{G}_{1}+\mathbf{G}_{2})/2\right\} $}
with the two reciprocal superlattice vectors $\mathbf{G}_{1}$ and $\mathbf{G}_{2}$. 
Its onset
is determined by a critical concentration of electrons or holes around charge neutrality and is signaled
by a significant distortion of the single-particle spectrum as depicted
Fig.~\ref{fig:weak-coupling-results}:
For a large amount of electron- or hole-doping, no electronic
symmetry breaking order develops (recall, we do not probe for superconductivity) 
and the characteristic van Hove singularities as well as the linear
dispersion relation near charge neutrality in the density of states are observed. 
When tuning the TBG system towards the magic-angle,
the lower (upper) critical band filling decreases (increases) and
the parameter regime of the stripe charge density wave order increases, as depicted in Fig.~\ref{fig:results}, 
due to an increasing $ \beta $. 

A real space representation of of the charge density wave order is depicted in the inset of Fig.~\ref{fig:weak-coupling-results}. It is noted that the corresponding real-space charge distribution, accessible, e.g., in STM measurements, differs qualitatively because of the highly non-local shape of Wannier functions with highest weight at the AA-stacked regions at the center of the hexagons. The local charge distribution would rather resemble a distorted version of the disordered state breaking $ C_3 $-rotational symmetry.

The observations are understood by setting up a corresponding mean field theory which is presented in the Appendix \ref{app:mean field-theory}. 
As charge modulations are predominantly determined by direct interaction processes with numerically large interaction matrix elements, exchange and pair-hopping processes are here neglected  
and the effective interaction in the corresponding channel is determined to 
\begin{equation}
 U_\text{CDW} = U_{\text{on-site}}+U_{\text{NN}}-4U_{\text{NNN}}-3U_{\text{NNNN}}, 
\end{equation}
which may be negative for sufficiently large $ U_{\text{NNN}} $, $ U_{\text{NNNN}} $ and small $ U_{\text{on-site}} $, $ U_{\text{NN}} $ interaction matrix elements.
This is made plausible by inspecting a possible real space representation of the charge density wave order depicted in the inset of Fig.~\ref{fig:weak-coupling-results}: On mean field level, this charge configuration minimizes interaction contributions from NNN and NNNN direct interaction processes. 
In particular for the numerical values of interaction matrix elements determined in Sec.~\ref{subsec:interaction-ME}, $  U_{\text{CDW}} / U_\text{on-site}\approx -2.18 $ yielding an effective attractive interaction strength. 
This finding is complemented by the result for the static charge susceptibility with finite momentum transfer $ \textbf{Q} $. 
It is peaked for doping levels around the CNP with small peaks at the van Hove points, but does not diverge due to the absence of a nesting condition (see the Appendix \ref{app:mean field-theory} for details). 
An onset of this order therefore requires a finite, attractive interaction strength 
and follows the qualitative illustration depicted in Fig.~\ref{fig:results}. 

Our results have to be contrasted to other types of charge density wave orders which rely on certain nesting conditions between the van Hove  points of the single-particle spectrum and which were discussed, e.g., in Refs.~\cite{Isobe18:RG,Laksono18:SCchargeOrder,Kozii19:NematicSCDensityFluct}. However, we do not find any evidence for the presence of this kind of instability within our modelling approach. 

\subsection{Strong-coupling regime}
\label{sec:strong-coupling-regime}

In the magic-angle regime, kinetic
energy contributions of electrons are expected to be much smaller
than contribution from interaction processes as $ \beta \gg 1 $.
To analyze possible electronic ground states, we therefore consider \textit{density-density interaction} processes
only to obtain an analytical tractable model. 
In this limit of "infinite couplings",
the Hamiltonian contains only contributions from direct and exchange
interaction processes and is given by 
\begin{multline}
\label{eq:HSC}
H_{\text{SC}}= \\ 
\tfrac{1}{2}\sum_{\sigma\sigma'}\sum_{ab\in\hexagon}(U_{ab}-J_{ab}\delta_{\sigma\sigma'}) (\hat{n}_{a\sigma}-\tfrac{1}{2})(\hat{n}_{b\sigma'}-\tfrac{1}{2}),
\end{multline}
where $\hat{n}_{\alpha\sigma}$ represents the local occupation number operator. 
As $ [ H_{\text{SC}}, \hat{n}_{a\sigma} ] = 0 $, this approximation renders the local occupation number a "good" quantum number and the theory classical. Later, kinetic contribution
may be incorporated perturbatively in orders of $\sim t/U$, which
is however not part of this work. 

The representation Eq.~\eqref{eq:HSC} of this model is particle-hole symmetric, i.e.~invariant under
$\hat{n}_{a\sigma}\leftrightarrow1-\hat{n}_{a\sigma}$, which allows us to study either hole or electron doping. 
As the theory is classical, the electronic ground state is determined by minimizing the energy functional associated with $ H_\SC $
with the local occupation
numbers as variational parameters. 
The optimization problem is solved
by using the Monte Carlo-based simulated annealing algorithm
\citep{Kirkpatrick83:SimulatedAnnealing}. 
Details about the employed procedure to determine electronic ground states are given in the Appendix \ref{app:sc}. 

\begin{figure}[t]
	\centering{}\includegraphics[draft=false,width=0.9\columnwidth]{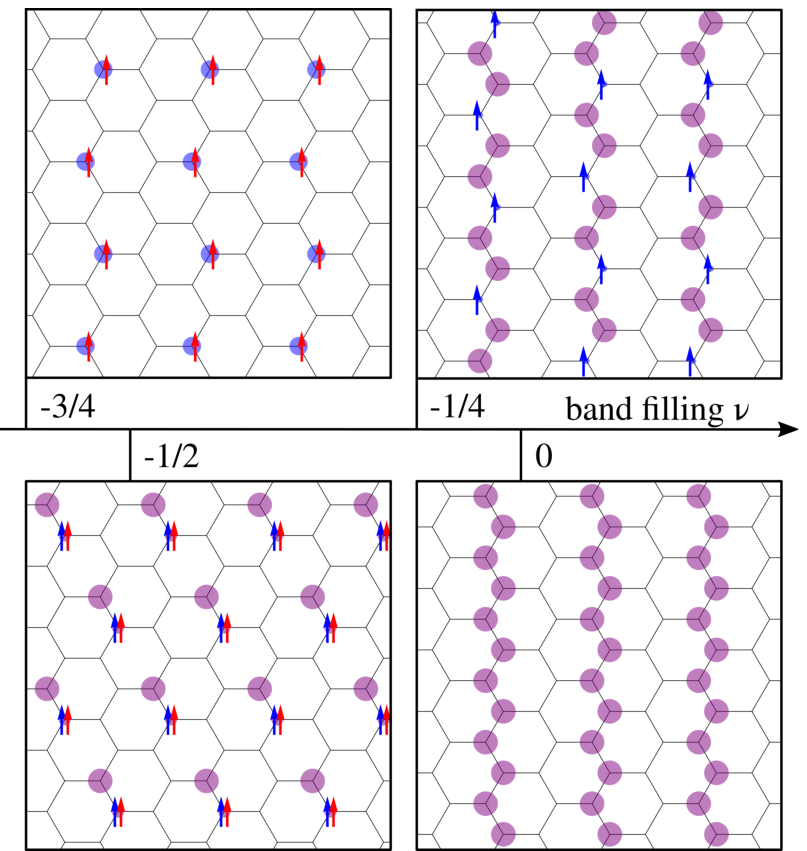}\caption{\label{fig:sc-results}
		Obtained charge configurations representing the electronic ground
		states in the \textit{strong-coupling regime} which exhibit Mott insulting
		behavior.
		The occupation number at one particular lattice site, which
		hosts in total 4 electronic states, is symbolically indicated as follows
		(arrows represent spin up/down states, colors red/blue valley $\xi=\pm$
		states): $4/4$ occupation %
		\parbox[c][1\baselineskip]{1.1cm}{%
			\protect\includegraphics[height=1.5\uppercaseHeight]{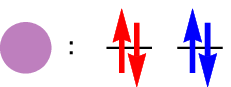}%
		}, $3/4$ occupation %
		\parbox[c][1\baselineskip]{1.1cm}{%
			\protect\includegraphics[height=1.5\uppercaseHeight]{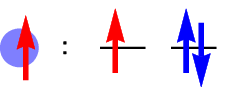}%
		} which is valley- and spin-polarized, $1/2$ occupation %
		\parbox[c][1\baselineskip]{1.1cm}{%
			\protect\includegraphics[height=1.5\uppercaseHeight]{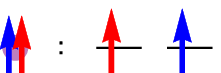}%
		} which is spin-polarized, $1/4$ occupation %
		\parbox[c][1\baselineskip]{1.1cm}{%
			\protect\includegraphics[height=1.5\uppercaseHeight]{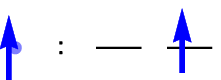}%
		} which is valley- and spin-polarized, else empty.}
\end{figure}

For the commensurate band fillings $\nu=0,\pm1/4,\pm1/2,\pm3/4$, 
we find Mott-insulating ground states which break different combinations of discrete translational, spin and (or) valley symmetries. 
When adding or removing electrons, i.e.~away from these commensurate fillings, 
we expect that the insulators turn into conductors where single particles
move in a landscape of potential barriers generated by electrons and
holes constituting the nearest Mott state. 
The obtained results for the ground state charge configurations for commensurate moiré band fillings of the hole-doped side are depicted in Fig.~\ref{fig:sc-results}. 
Note that because the system exhibits a spontaneous symmetry breaking, the depicted configurations are only particular realizations out of several possible ground states all characterized by the same set of broken symmetries, respectively. 

For moir\'e band fillings $\nu = 0,\pm1/4$, we find stripe-type orders
which resemble our findings of the weak-coupling approach with
charge inhomogeneities described by ordering vectors given
in Eq.~(\ref{eq:orderPar}). It indicates that this particular density
configuration minimizes the potential energy costs generated by the
dominant direct interaction processes irrespective of kinetic energy
contributions which is in line with our previous finding that the formation
of density inhomogeneities in the weak coupling-regime is not linked
to features of the single-particle spectrum. 
For $\nu=\pm1/2,\pm3/4$,
we find charge configurations which maximize the distance between
charges similar to the principal of Wigner crystallization. 

We conclude that the charge distribution is decisively determined
by direct interaction processes $U$ which are characterized by a
significant coupling of all sites belonging to one hexagon of the superlattice. It is determined
solely by the ratio of direct interaction matrix elements which was determined
to $(U_{\text{on-site}}\negmedspace:\negmedspace U_{\text{NN}}\negmedspace:\negmedspace U_{\text{NNN}}\negmedspace:\negmedspace U_{\text{NNNN}})/U_{\text{on-site}}=(1\negmedspace:\negmedspace0.79\negmedspace:\negmedspace0.63\negmedspace:\negmedspace0.58)$,
where the exact numerical values matter as, e.g., simple ratios of type
$(1\negmedspace:\negmedspace\frac{2}{3}\negmedspace:\negmedspace\frac{1}{3}\negmedspace:\negmedspace\frac{1}{3})$, which are connected to the amount of direct overlap of wannier functions, lead to different results. 
Furthermore, since the local
single-particle states are either empty or occupied, the particular
ground state is required, unless occupied lattice sites are always
fully occupied, to additionally break the spin- and/or the valley-symmetry.
Since direct interaction processes do not discriminate between spin
and charge degrees-of-freedom, the energetically most favorable configuration
is here determined by the exchange interactions. 
Their matrix elements,
for both the intra- and intervalley channel, are always found to
be positive and therefore favor an alignment of spins non-locally
(because of intravalley exchange) and locally (because of on-site intervalley exchange). 
This results for $\nu=\pm1/4,\pm3/4$ in a condensation of
local degrees of freedom of partially occupied sites in one particular
spin and valley sector, whereas for $\nu=\pm1/2$ in one particular
spin sector. 

Our findings have to be contrasted to similar strong coupling approaches presented in Refs.~\citep{Kang19:StrongCouplingPhase,Xu18:QMCKekule}. In Ref.~\citep{Kang19:StrongCouplingPhase}, the authors assume an averaged interaction strength for all 
processes connecting the localized states of one hexagon and also included processes beyond the density channel. 
In Ref.~\cite{Xu18:QMCKekule}, the ground state analysis is conducted for direct interactions processes only with a fixed ratio of interaction matrix elements connected to the amount of direct overlap of neighboring Wannier functions. 
As our results depend decisively on the distance dependence of interaction elements, our ground states for commensurate band fillings $\nu=\pm1/2,\pm3/4$ differ. 

\section{Conclusion}
In this work, we found a hierarchy of interaction processes as specified in 
Eq.~(\ref{eq:hierarchy-ME}). 
Here, direct interaction processes dominate  
followed by intra- and intervalley
exchange interaction processes which are at least one order in magnitude
smaller but always positive. 
These interaction processes are necessary to determine the electronic ground state unequivocally. 
Combined with the distance dependence of the matrix elements which connect, to leading order, all Wannier states which belong to the same hexagon formed by the AB- and BA- stacked regions of the superlattice, these characteristics were found decisive for the determination
of possible electronic ground states. 

The most robust finding of our analysis, that occurs at weak and strong
coupling, is the emergence of a nematic state that breaks the three-fold
rotational symmetry of the moir\'e lattice. The details of the related
translational symmetry breaking and of additional broken symmetries
depend then on the strength of the interactions and the filling fractions.
While critical fluctuations, not included in our formalism, may render
charge density waves, spin, or valley order finite ranged, the discrete
nematic symmetry breaking should give rise to a sharp second-order
phase transition with a finite transition temperature. Even for a moderate
symmetry-breaking substrate-induced strain we expect a well-defined
crossover temperature. 

In addition to the nematic state, we find an
onset of spin- and valley-polarized orders at strong couplings. This
effect is due to the non-local, positive
intravalley and intervalley exchange couplings suggesting modified Hund's rule, where
first the spin and subsequently the valley number is maximized when filling up
superlattice sites with electrons. 
We expect that this finding is consistent with the degeneracy
pattern of the Landau levels of the insulating states observed in quantum oscillations. However, it deserves a more thorough investigation of this aspect to confirm this conclusion. 

If the nematic order exist away from commensurate band fillings, the reduced symmetry at $T_{c}$ excludes more complex superconducting order parameters,
such as chiral $d+id$ or nematic $d+id$ states. On the other hand,
the abundance of the nematic order in twisted bilayer suggests that nematic fluctuations may be important in inducing or amplifying superconductivity
in these materials as, e.g., discussed in Refs.~\citep{Lederer15:EnhancedSCNematicQCP,Lederer17:SCNeamticQCP,Kozii19:NematicSCDensityFluct}. 

\textit{Acknowledgements}: We thank B. Anderson, R. Fernandes, J.
Kang, L. Merkens, J. Schmalian and A. Wechselberger for insightful
discussions. 
The author, furthermore, acknowledges support by the KIT-Publication Fund of the Karlsruhe Institute of Technology.

\bibliography{references}

\appendix

\begin{widetext}
	
\section{\label{sec:Construction-of-wannier}Construction of maximally localized	Wannier functions }

\begin{figure}[b]
	\subfloat[Wannier function of sublattice state $\alpha=BA$.]{\centering{}\includegraphics[width=1\linewidth]{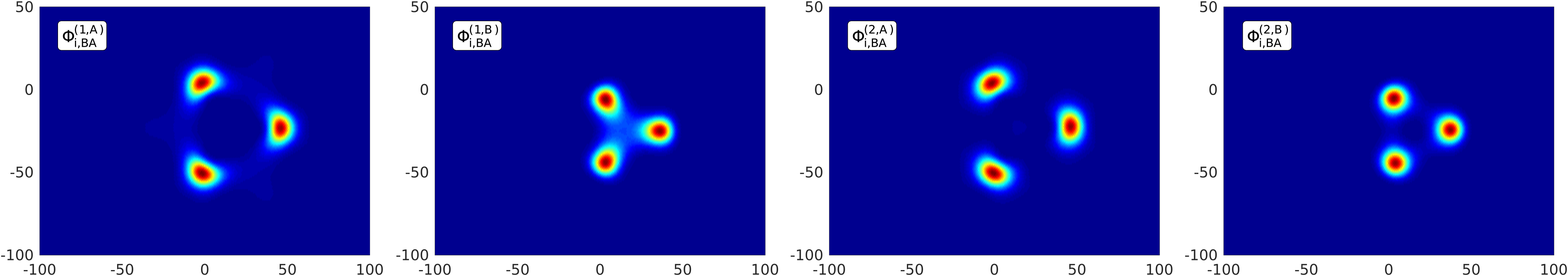}}	
	\LyXZeroWidthSpace\subfloat[Wannier function of sublattice state $\alpha=AB$.]{\centering{}\includegraphics[width=1\linewidth]{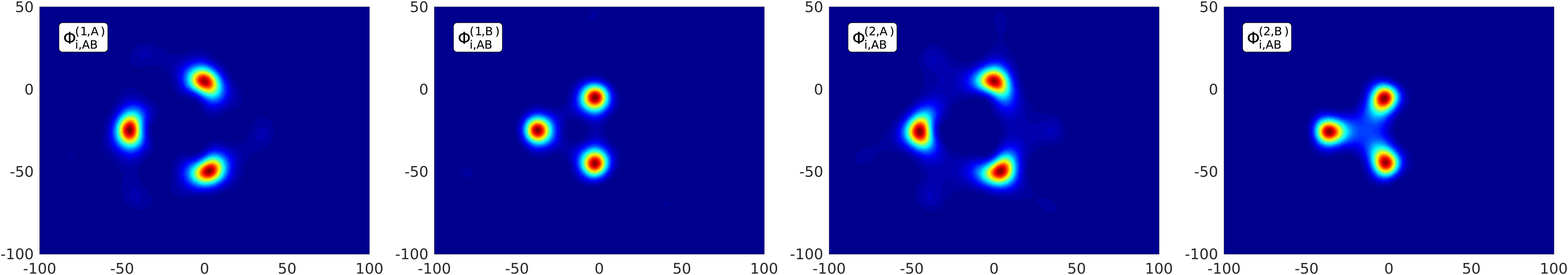}}
	\caption{\label{fig:orbitals}
		Absolute amplitude of the constructed Wannier functions $\Psi_{i\alpha} = \sum_{j\gamma}\Phi_{i\alpha}^{(j,\gamma)}  $
		for $\theta=1.05\lyxmathsym{\protect\textdegree}$ projected on the single-layer
		graphene sites labelled by the graphene layer index $j\in \{1,2\}$ and the graphene crystalline sublattice index.} 
\end{figure}

The Wannier basis is constructed following the method of maximally
localized Wannier functions \citep{Marzari97:MaxLocWanFunc,Marzari12:ReviewMaxLocWannierFunctions}.
In the case of twisted bilayer graphene, we follow the approach presented
in Ref.~\citep{Koshino18:LiangMaxLocWO} where the valley degrees of freedom
are assumed to fully decouple in the limit of small twist-angles.
We expect that this approach is equivalent to other two-orbital approaches
\citep{Po18:SenthilTheoryPaper,Kang18:LocWannierStates} which drop
the requirements for a valley symmetry at first hand, but recover
an approximate valley symmetry later.
As the constructed localized states
possess a definite valley number, we drop the valley $\xi$ and the spin $\sigma$ quantum numbers from the subsequent analysis while determining the wave functions for one particular valley. 
The wave functions of the other valley are obtained by complex conjugation \cite{Koshino18:LiangMaxLocWO,Po18:SenthilTheoryPaper}.

Represented in real space as projections on the two graphene layers, the Bloch functions of the corresponding moir\'e bands introduced in Eq.~\eqref{eq:Hm} are given by 
\begin{equation}
\label{eq:A1}
\psi_{\lambda\mathbf{k}}\left(\mathbf{r}\right)=\psi_{\lambda\mathbf{k}}^{\left(1,A\right)}(\mathbf{r})+\psi_{\lambda\mathbf{k}}^{\left(1,B\right)}(\mathbf{r})+\psi_{\lambda\mathbf{k}}^{\left(2,A\right)}(\mathbf{r})+\psi_{\lambda\mathbf{k}}^{\left(2,B\right)}(\mathbf{r}),
\end{equation}
with the crystal momentum $ \mathbf{k} $, which is element of the moir\'e Brillouin zone. The projection on the graphene layer $j\in \{1,2\}$ and the graphene crystalline sublattice $\gamma \in \{A,B\}$ is given by 
\begin{eqnarray}
\label{eq:A2}
\psi_{\lambda\mathbf{k}}^{\left(j,\gamma\right)}(\mathbf{r}) & = & N^{-1/2}\sum_{im}U_{jm}^{(\lambda)}(\mathbf{k})e^{i(\mathbf{k}+\mathbf{G}_{m})\mathbf{a}_{i}^{\left(j\right)}}\phi(\mathbf{r}-\mathbf{a}_{i}^{\left(j\right)}-\mathbf{u}_{\gamma}^{\left(j\right)}),
\end{eqnarray}
where the unitary matrix $ U_{jm}^{(\lambda)} (\textbf{k}) $ connects the moir\'e Bloch state labelled by $ \mathbf{k} $ and the moir\'e band index $ \lambda $, and the graphene tight-binding basis which is obtained by diagonalizing Eq.~\eqref{eq:Htbg}. 
$\mathbf{G}_{m}$ denotes a reciprocal superlattice vector, 
whereas the Bravais lattice vector of the graphene layer $ j $ is represented by $\mathbf{a}_{i}^{\left(j\right)}$ and the crystalline basis vector by $\mathbf{u}_{\gamma}^{\left(j\right)}$. 
$\phi(\mathbf{r})$ represents graphene $p_{z}$-orbitals localized at
$\mathbf{r}=0$. 

Within the method of maximally localized Wannier functions, Wannier functions are given by a linear superposition of Bloch wave functions weighted by an exponential phase factor \citep{Marzari97:MaxLocWanFunc,Marzari12:ReviewMaxLocWannierFunctions}. 
Here, the Wannier function, which is located in superlattice unit cell $i$ and centered at the high symmetry points $\alpha \in \{\text{AB,BA}\}$ identified with the AB- and BA-stacked
regions of the superlattice, is given by 
\begin{equation}
\label{eq:A3}
\Psi_{i\alpha}\left(\mathbf{r}\right)=N^{-1/2}\sum_{\lambda\mathbf{k}}e^{-i\mathbf{k}\mathbf{A}_{i}}\mathcal{U}_{\lambda\mathbf{k}}^{\left(\alpha\right)}\psi_{\lambda\mathbf{k}}\left(\mathbf{r}\right)
\end{equation}
with the superlattice vector $\mathbf{A}_{i}$. 
To obtain maximally localized Wannier function, the unitary matrix
$\mathcal{U}_{\lambda\mathbf{k}}^{\left(\alpha\right)}$ is chosen
such that the spread functional
\begin{equation}
g[\mathcal{U}]\equiv \int d^{d}r\,\Psi_{i\alpha}^{*}\left(\mathbf{r}\right)(\mathbf{r}-\mathbf{R}_{i\alpha})^{2}\Psi_{i\alpha}\left(\mathbf{r}\right)
\end{equation}
is minimal. Here, $\mathbf{R}_{i\alpha}$ represents the coordinates
of the Wannier function's center located at the center of the AB- or
BA-stacked regions of the $i$th superlattice unit cell.

By following Ref.~\cite{Koshino18:LiangMaxLocWO} in choosing the initial guess for $\mathcal{U}_{\lambda\mathbf{k}}^{\left(\alpha\right)}$, 
the optimal unitary matrix is obtained by employing multidimensional optimization procedures. 
As an example, the obtained Wannier functions, 
which are checked to be exponentially localized, 
for a twist-angle of $\theta=1.05\lyxmathsym{\textdegree}$ and a particular valley,
are depicted in Fig.~\ref{fig:orbitals} 
as projections on the single layer graphene states
obtained by rearranging Eqs.~\eqref{eq:A1}-\eqref{eq:A3}.
Having established the single-particle Wannier basis whose real-space representation is given by the Wannier functions centered at the corresponding high symmetry points of the superlattice, the single-particle transition amplitudes and the interaction matrix elements of the two-particle interaction processes between Wannier states are computed straightforwardly as discussed in the next two subsections. 

\subsection{Single-particle transition amplitudes}
\label{subsec:hopping-parameter}

\begin{figure}[t]
	\begin{centering}
		\includegraphics[width=0.5\linewidth]{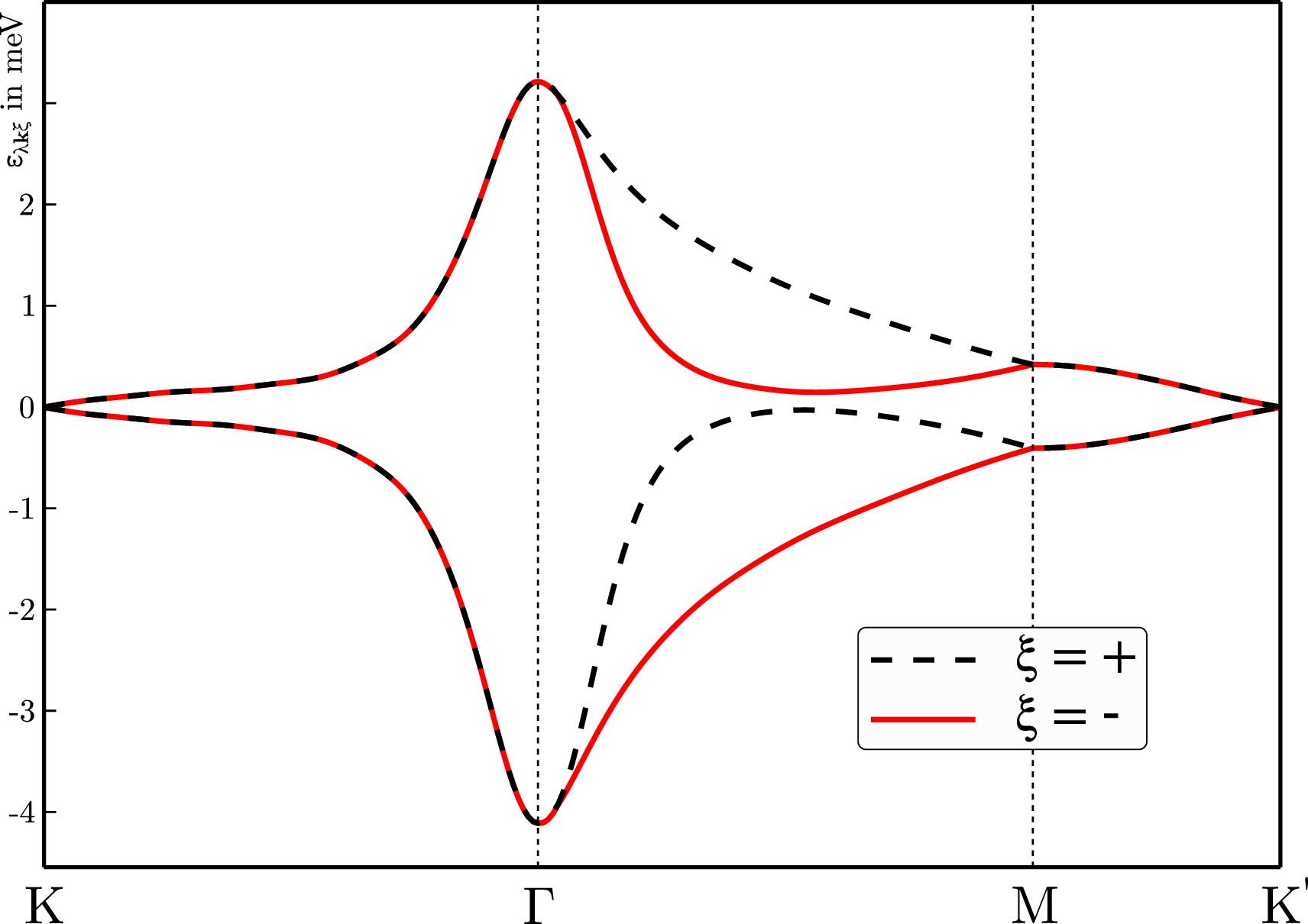}\caption{\label{fig:Moire-bands}Single-particle moir\'e band spectrum which is obtained
			by diagonalizing the effective tight-binding model Eq.~(\ref{eq:eff-model})
			for a twist-angle $\theta=1.05\lyxmathsym{\protect\textdegree}$. The dashed-black and red line depict the bands with
			valley number $\xi=+$ and $\xi=-$, respectively. }
		\par\end{centering}
\end{figure}

The single-particle transition amplitudes,
which enter the effective tight-binding 
model introduced in Eq.~(\ref{eq:eff-model}) and which are by construction diagonal
in valley and spin space, 
are computed by applying the inverse unitary transformation determined previously to the free Hamiltonian specified in Eq.~\eqref{eq:Hm}. Hence, hopping parameters are given by 
\begin{align}
t_{i\alpha,j\beta} & =N^{-1}\sum_{\mathbf{k}\lambda}e^{i\mathbf{k}(\mathbf{A}_{i}-\mathbf{A}_{j})}\mathcal{U}_{\lambda\mathbf{k}}^{(\alpha)\dagger}\epsilon_{\lambda\mathbf{k}}\mathcal{U}_{\lambda\mathbf{k}}^{(\beta)}
\end{align}
with the single-particle energy $ \epsilon_{\lambda \mathbf{k}} $. 

We observe that the amplitude of transition amplitudes drops rather
slowly with distance: To recover the weakly dispersing moir\'e bands
of the Hamiltonian introduced in Eq.~(\ref{eq:Hm}),
we have to take transition amplitudes between orbitals with a spatial
separation of more than 10 superlattice unit cells into account. The
single-particle moir\'e spectrum which is obtained by means of the determined transition amplitudes $\{t_{i\alpha,j\beta}\}$
is depicted in Fig.~\ref{fig:Moire-bands} and matches the spectrum which was previously determined by employing the continuum model yielding Eq.~\eqref{eq:Hm}.

\subsection{Interaction matrix elements}
\label{subsec:interaction-matrix-elements}

\begin{figure}[b]
	\hfill
	\subfloat[\label{fig:ME-a}Interaction matrix elements, $ U_\text{on-site} $, $ U_\text{NN} $, $ U_\text{NNN} $, $ U_\text{NNNN} $, $ J_\text{NN} $ and $ X_\text{NN} $ as function of twist-angle $ \theta $ in units of $ e^2/\epsilon L_M $. The other matrix elements, which are smaller and are not depicted here, scale equilvalently. 
	]{\includegraphics[width=0.49\linewidth]{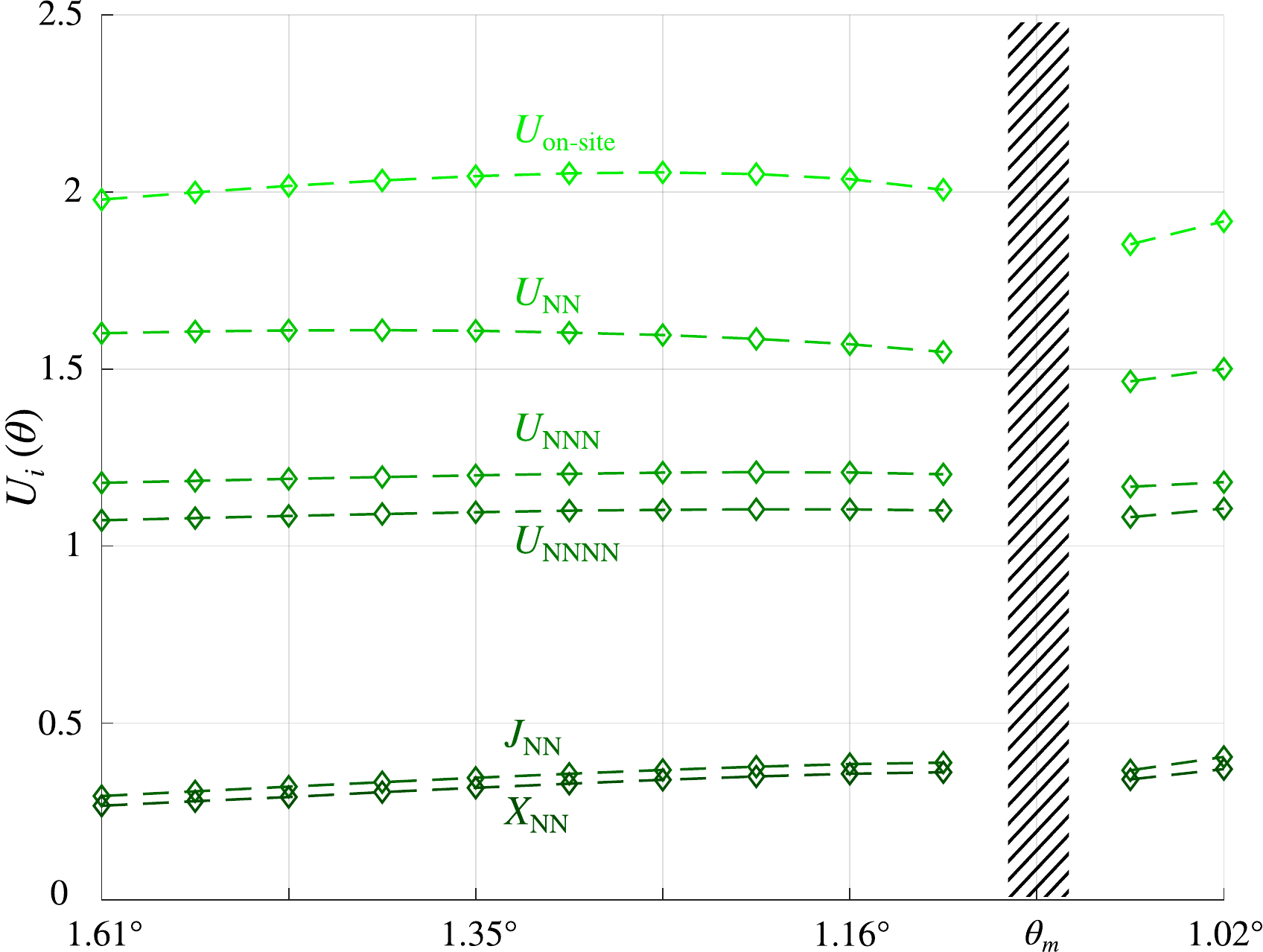}}
	\hfill
	\subfloat[\label{fig:ME-b}Ratio of screened and unscreend interaction matrix elements $ U_i^{\text{scr}} / U_i $ for $ U_\text{on-site} $, $ U_\text{NN} $, $ U_\text{NNN} $ and $ U_\text{NNNN} $ as function of twist-angle $ \theta $. 
	The screening length is chosen to $ \xi_{\text{scr}} = 80 \, a \approx 19.4 \text{nm} $. ]{
		\includegraphics[width=0.49\linewidth]{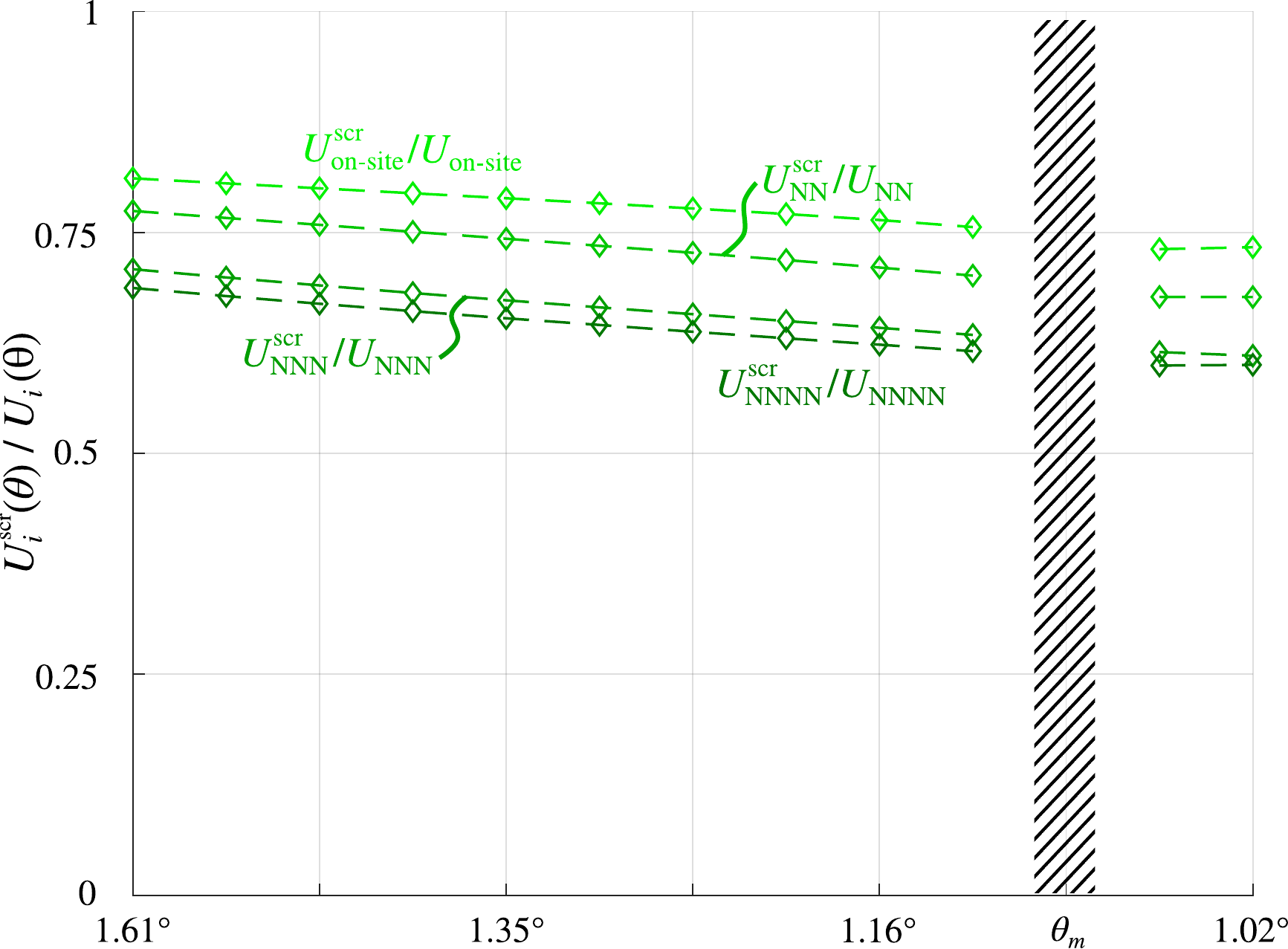}}
	\hfill
	\caption{\label{fig:intME-twistangle}Interaction matrix elements as function of the twist-angle and in the presence of screening due to a finite distance of the TBG system to the metallic back gate. }
\end{figure}

The interaction matrix elements between Wannier states which enter the interacting tight-binding model introduced in Eq.~\eqref{eq:eff-model} are determined by evaluating the expression 
\begin{equation}
U^{(\text{scr})}_{abcd}= \int_{{\bf r}{\bf r}'} 
\Psi_{a\sigma}^{\dagger}\left(\mathbf{r}\right)\Psi_{b\sigma'}^{\dagger}\left(\mathbf{r}'\right)
V_{(\text{scr})}(\mathbf{r}-\mathbf{r}') \Psi_{c\sigma'}\left(\mathbf{r}'\right)\Psi_{d\sigma}\left(\mathbf{r}\right), 
\end{equation}
where the Wannier function $ \Psi_{a\sigma}(\textbf{r}) $ represents the single-particle wave function of the Wannier state $ a = (i,\alpha,\xi) $ with spin $ \sigma $. 
The interaction potential is chosen first to an unscreened Coulomb potential $ V(\textbf{r}) = \frac{e^{2}}{4\pi\epsilon} \frac{1}{|\textbf{r}|} $. 
Second, the effect of screening is investigated by considering a Yukawa-type potential $ V_\text{scr}(\textbf{r}) = \frac{e^{2}}{4\pi\epsilon} \frac{e^{-|\textbf{r}|/\xi_{\text{scr}}}}{|\textbf{r}|}  $ which is parametrized by the screening length $ \xi_\text{scr} $. 
As the distance of the TBG sample to the metallic back gate is determined by the thickness of the hBN layer, which ranges between $ 10\dots30\text{nm} $ \citep{Cao18:PabloExpSuperconductivity,Cao18:PabloExpInsulatorAtHalfFill}, screening effects are expected to be relevant for superlattice unit cell sizes of order of this  distance. 
The screening length is therefore chosen to $ \xi_{scr} = 80 \, a \approx 19.4 \text{nm} $. 
The results for the interaction matrix elements as function of the twist-angle and under the effect of screening, which summarize our general findings, are depicted in Fig.~\ref{fig:intME-twistangle}. 
Our main finding is that the twist-angle dependence of the relative strength of interactions is described, in leading order, by the dimensionless constant $ \beta = \frac{e^2}{\epsilon L_M \Lambda} $ where the changes of the interaction matrix elements due to variations of the twist-angle is determined by the superlattice constant $ L_M \propto 1/\sin(\theta/2) $, 
and that the ratios between the various interactions elements given in Fig.~\ref{fig:interaction-ME} are representative for a range of twist-angles near the magic-angle regime. 
Further changes due to twist-angle variations or screening effects are subleading and do not affect the ground state analysis presented in Sec.~\ref{sec:ground-state-analysis}. 

The twist-angle dependence of the interaction matrix elements is depicted in Fig.~\ref{fig:ME-a}. 
Their dependence is rather weak when expressed in units of $ e^2/\epsilon L_M $ and can be safely neglected in the present work. 
The effect of screening on the amplitude of interaction matrix elements is depicted in Fig.~\ref{fig:ME-b}. 
We observe an overall change in the amplitude which generally reduces the strength of interactions effects. 
This effect can be compensated in a redefinition of $ \beta $.
Furthermore, it is found that the amplitude of non-local interaction processes is stronger suppressed than for local interaction processes which is as expected. However, as the neighboring Wannier functions have still significant overlap, this effect is found minor, at least for interaction processes which connect Wannier state belonging to one hexagon of the superlattice which are found relevant for the present work. 
In particular, the interaction strength in the stripe charge density wave channel is reduced from $  U_{\text{CDW}} / U_\text{on-site}\approx -2.18 $ to $ U^{\text{scr}}_{\text{CDW}} / U_\text{on-site}^\text{scr} \approx -1.58 $. 
Furthermore, the ratio between local- to non-local processes changes from $(U_{\text{on-site}}\negmedspace:\negmedspace U_{\text{NN}}\negmedspace:\negmedspace U_{\text{NNN}}\negmedspace:\negmedspace U_{\text{NNNN}})/U_{\text{on-site}}=(1\negmedspace:\negmedspace0.79\negmedspace:\negmedspace0.63\negmedspace:\negmedspace0.58)$ 
to $ (U^{\text{scr}}_{\text{on-site}}\negmedspace:\negmedspace U^{\text{scr}}_{\text{NN}}\negmedspace:\negmedspace U^{\text{scr}}_{\text{NNN}}\negmedspace:\negmedspace U^{\text{scr}}_{\text{NNNN}})/U^{\text{scr}}_{\text{on-site}} = (1\negmedspace :\negmedspace 0.77\negmedspace :\negmedspace 0.52\negmedspace :\negmedspace 0.46) $. 
However, it is found that the obtained results for the weak and strong-coupling regime presented in Sec.~\ref{sec:weak-coupling-regime} and \ref{sec:strong-coupling-regime} are robust against these changes. 

\section{Numerical ground state analysis in the weak-coupling regime\label{sec:Lanczos-algorithm}}

To identify the electronic ground state in the weak-coupling regime, a  mean field analysis is conducted where all relevant interaction terms are decoupled locally by introducing local mean fields yielding the mean field Hamiltonian $ H_{\text{MF}} $ introduced in Eq.~\eqref{eq:MF-ham}. 
The local mean fields are variational parameters which have to be determined self-consistently. 
The ground state is eventually obtained by minimizing
an energy functional which derives from the mean field Hamiltonian. 
The minimization procedures is conducted numerically 
on a finite lattice of $30\times30$ superlattice unit cells to capture
the rather slowly decaying transition amplitudes $\{t_{ab}\}$. 
Mutually
independent mean fields are introduced for a lattice of $6\times6$
superlattice unit cells with imposed periodic boundary conditions to capture possible electron orders which break translational symmetries. 
The algorithm to determine the electronic ground is presented in the following. 

\subsection{Numerical procedure}
The quadratic mean field Hamiltonian is given by (the spin
index is dropped for the matter of representation) 
\begin{equation}
H_{\text{MF}}=\sum_{ab}h_{ab}c_{a}^{\dagger}c_{b}, 
\end{equation}
where the matrix elements $ h_{ab} = h_{ab} [\langle c^\dagger c \rangle ] $ are given in the mean field Hamiltonian introduced in Eq.~\eqref{eq:MF-ham} and contain the mean fields $ \langle c^\dagger_a c_b \rangle $. 
In the present problem, $ h= (h_{ab}) $ is a hermitian $d\times d$
matrix  where
$d$ represents the dimensionality of the Hilbert space, which is $d=7200$ for the introduced finite lattice. The energy functional is obtained as the thermal expectation
value of the mean field Hamiltonian, 
\begin{equation}
E[\langle c^{\dagger}c\rangle]=Z^{-1}\text{tr}[e^{-\tfrac{H_{\text{MF}}}{k_B T}}H_{\text{MF}}] \label{eq:energyFunc} = \langle  H_{\text{MF}} \rangle , 
\end{equation}
which is evaluated in the zero temperature  limit. For this, correlations of type $\langle c_{a}^{\dagger}c_{b}\rangle$ have to be determined self-consistently under the condition to minimize $ E $. 

To compute the mean fields for a given mean field configuration, we determine the single-particle Green's function  
\begin{equation}
G_{ab}^{R/A}(\omega)=\left[\omega-h[ \langle c^\dagger c \rangle ]\pm i0^{+}\right]_{ab}^{-1} \label{eq:mean field-GF}
\end{equation}
which is again connected to the mean fields by 
\begin{equation}
\langle c_{a}^{\dagger}c_{b}\rangle=i\int_{-\infty}^{\mu}\frac{d\omega}{2\pi}\left[G_{ab}^{R}\left(\omega\right)-G_{ab}^{A}\left(\omega\right)\right].\label{eq:corrs}
\end{equation}
This self-consistent set of equations represents a certain gap equation which is susceptible to various electronic orders whose commensurability is set by the boundary conditions. 

In what follows, this gap equation is solved numerically. The determination of the inverse of a large matrix as required in Eq.~\eqref{eq:mean field-GF} is very costly. To overcome
this difficulty, we locally approximate the matrix inversion around
a certain state $a$ employing the Lanczos algorithm \citep{Lanczos50}.
This approximation procedure is applicable because $h$ possesses
a local structure, i.e.~state $a$ is locally coupled to only a handful
of other states $b$. The dimensionality of the Lanczos space $d_{L}$
is therefore much smaller than the original Hilbert space dimensionality
$d\gg d_{L}$ but still approximates the Green's function  accurately. Good results are here obtained for $d_{L}=50$. 

To approximate $G_{ab}^{R/A}(\omega)$, we construct the Lanczos space
around an initial state $a$. The transformation is given by a $d\times d_{L}$
unitary matrix $u$ with $u_{i1}=\delta_{ia}$ such
that $\tilde{h}=u^{\dagger}hu$ represents a tridiagonal hermitian
matrix \citep{Golub13:MatrixComputations}. 
Because of its tridiagonal form and its reduced rank, the
propagator in the reduced Lanczos space is readily determined exactly to $\tilde{G}^{R/A}(\omega)=[\omega-\tilde{h}\pm i0^{+}]^{-1}$
and the single-particle Green's function is eventually given by 
\begin{equation}
G_{ab}^{R/A}(\omega)\approx[u\tilde{G}^{R/A}(\omega)u^{\dagger}]_{ab}.
\end{equation}

\subsection{Numerical analysis}
\begin{figure}[b]
	\subfloat[\label{fig:o1}Stripe order I]{\centering{}\includegraphics[draft=false,width=.3\linewidth]{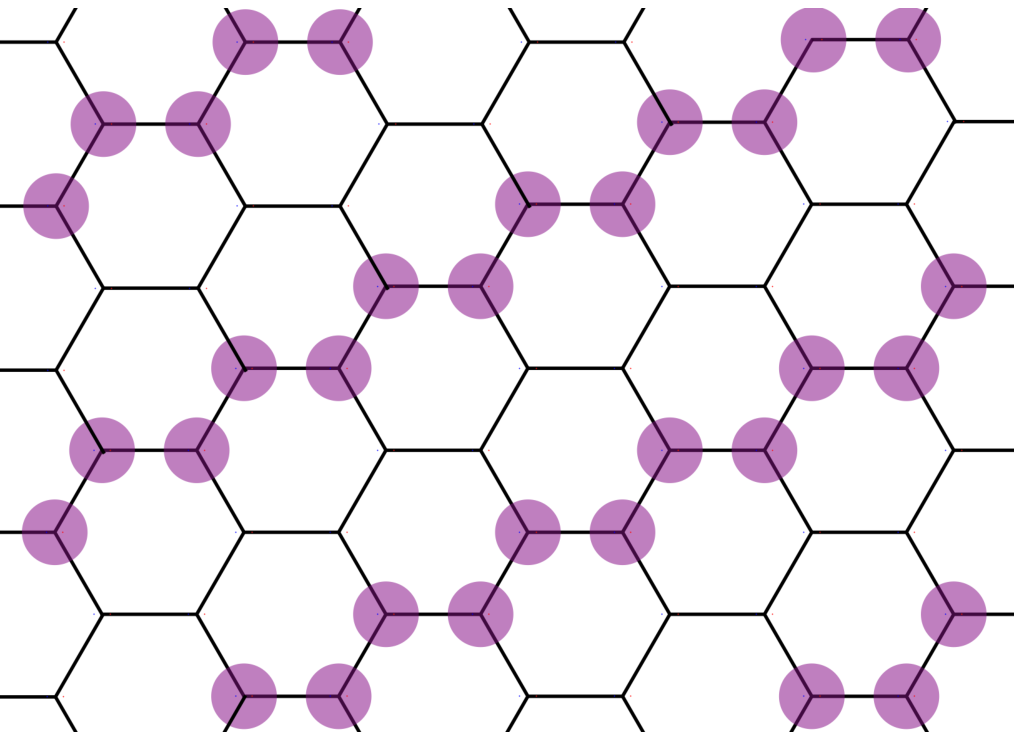}}
	\hfill \subfloat[\label{fig:o2}Stripe order II]{\centering{}\includegraphics[draft=false,width=.3\linewidth]{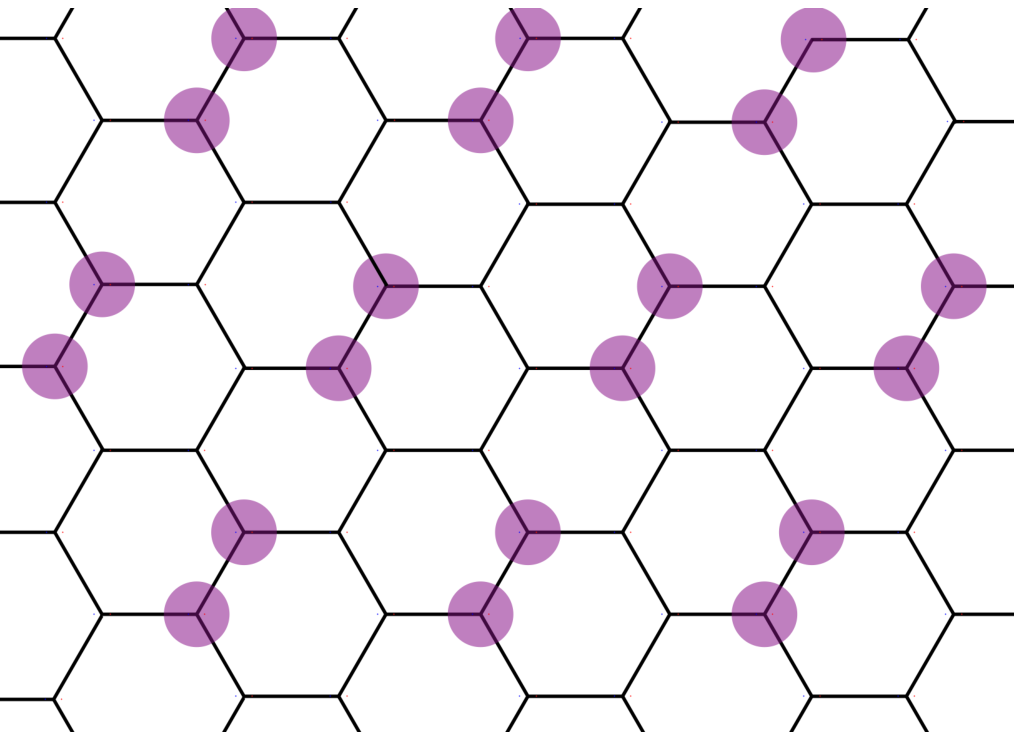}}
	\hfill \subfloat[Ferromagnetic stripe order II ]{\centering{}\includegraphics[draft=false,width=.3\linewidth]{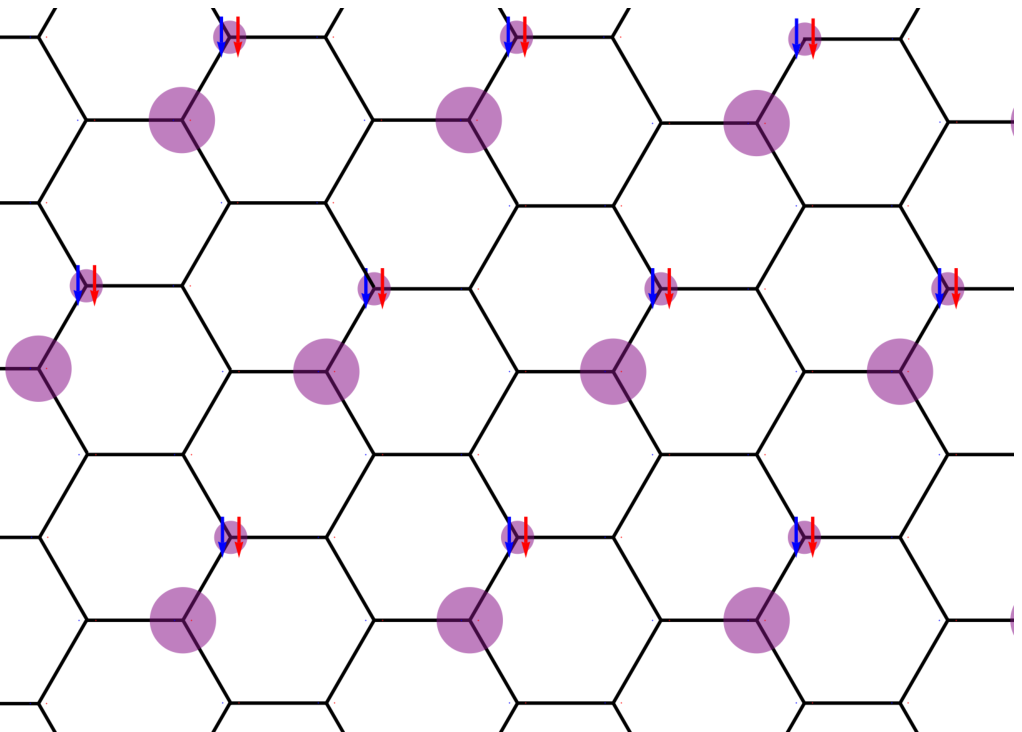}}
	\\ 
	\subfloat[Double stripe order]{\centering{}\includegraphics[draft=false,width=.3\linewidth]{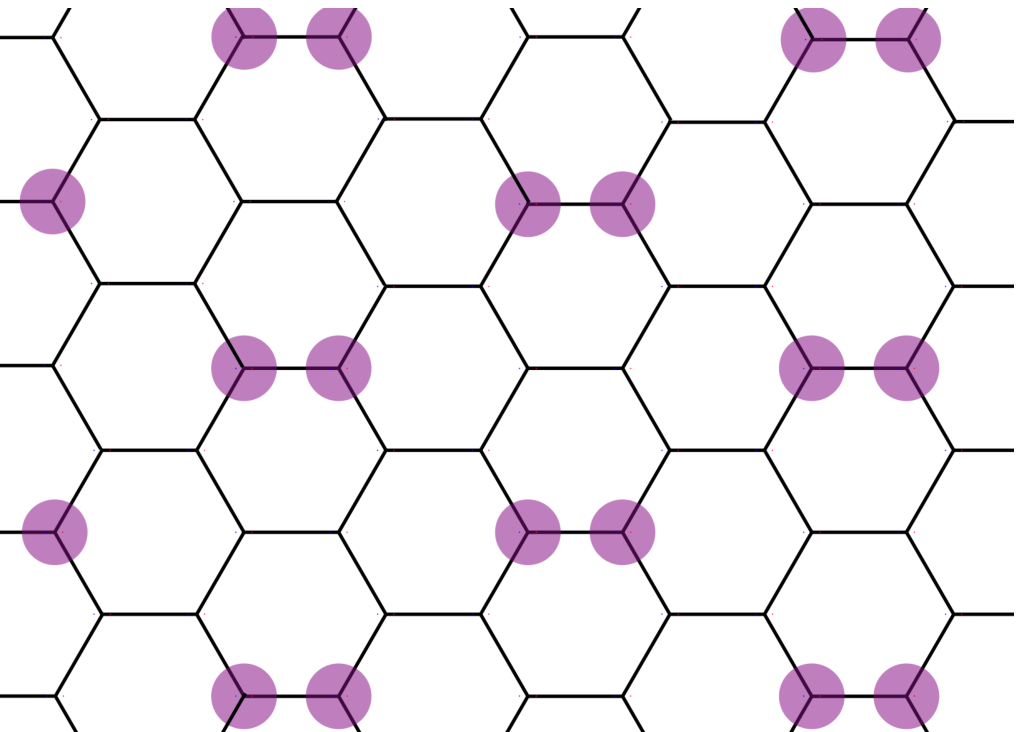}}
	\hfill \subfloat[Order with sublattice polarization, see e.g.~Ref.~\cite{Koshino18:LiangMaxLocWO}]{\centering{}\includegraphics[draft=false,width=.3\linewidth]{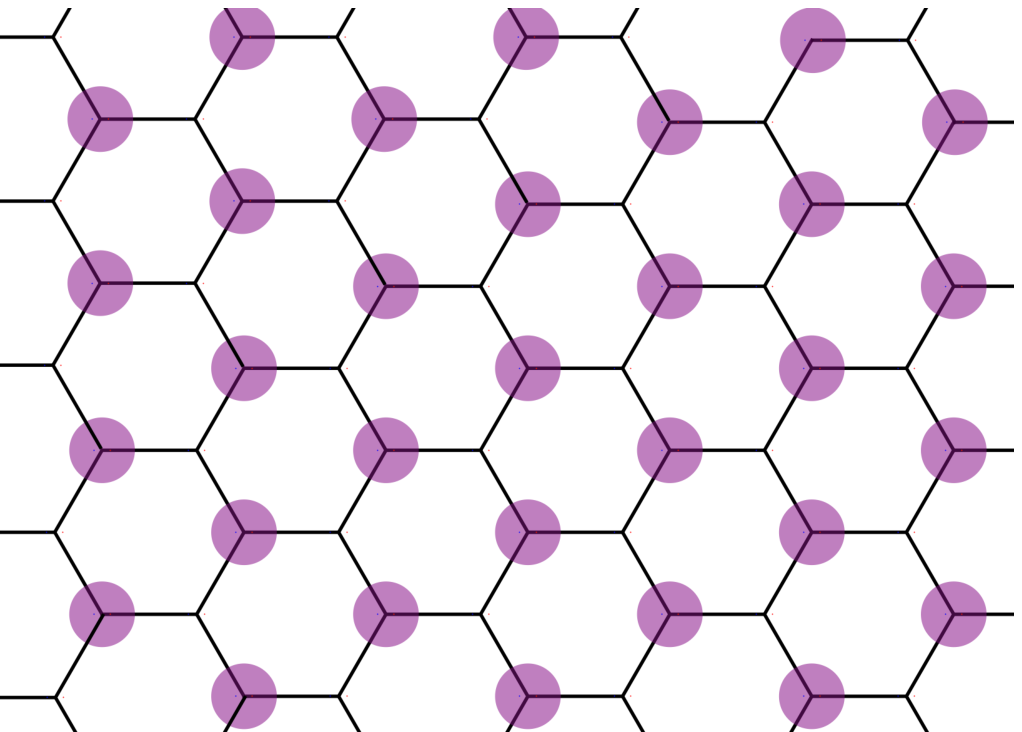}}
	\hfill \subfloat[Ferromagnetic order, see e.g.~Ref.~\cite{Venderbos19:CorrRafael,Seo19:FerroMott}]{\centering{}\includegraphics[draft=false,width=.3\linewidth]{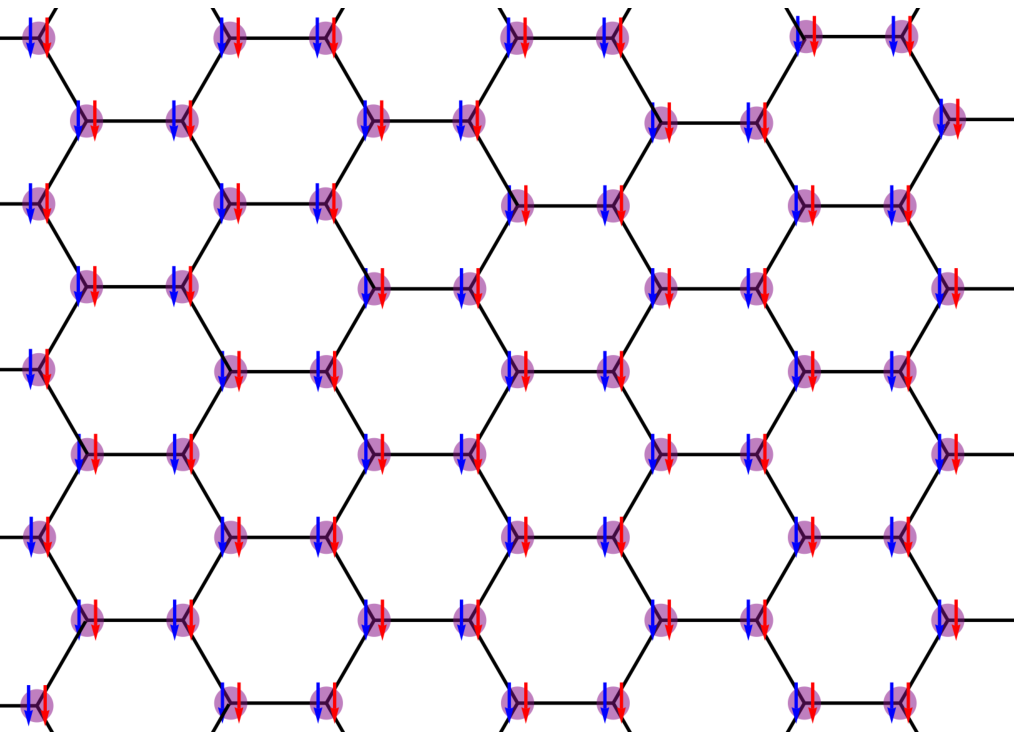}}
	\caption{\label{fig:order}Various electronic orders, which differ in the set of broken discrete symmetries. The purple dots and the up/down-arrow-pictograms represent the local densities and the spin polarizations as indicated in Fig.~\ref{fig:weak-coupling-results} and \ref{fig:sc-results}.	
		The stability and the associated energy of the depicted orders were checked explicitly to identify the true ground state. }
\end{figure}

To solve the coupled equations \eqref{eq:mean field-GF} and \eqref{eq:corrs}, we employ an iterative scheme where we start with a randomized initial mean field configuration and compute the resulting mean field configuration. This configuration serves as the initial configuration for the subsequent computational step. This sequence is repeated a finite number of times until a stable fixed point is reached. 
The energy associated with the state of the fixed point is found to minimize the energy functional Eq.~(\ref{eq:energyFunc}) and the corresponding mean field configuration, therefore, characterizes the electronic ground state of the system. 

To determine the interaction channel, which first develops an instability, we steadily increase the effective interaction strength $ \beta $ until a symmetry breaking electronic order develops. We find for  $ \beta \gtrsim 0.04 $ the symmetry breaking order depicted in Fig.~\ref{fig:o1}. This stripe charge density wave order is specified by the order parameter given in Eq.~\eqref{eq:orderPar}. 

To investigate the robustness of this result, 
we also analyze the stability of other electronic orders. 
For this, we restrict the gap equation Eq.~\eqref{eq:corrs} to the order parameters of the orders depicted in Fig.~\ref{fig:order}, respectively, and investigate their stability. Consequently, we compare their energies to identify the true ground state.  
The choice of the candidates depicted in Fig.~\ref{fig:order} is either motivated by our findings of the strong coupling regime Sec.~\ref{sec:strong-coupling-regime}, or by electronic orders discussed in the literature (see the captions for details). 

For the range of interaction strengths $0.1 \geq \beta \gtrsim 0.04 $, we find only two types of stripe charge density wave orders, Figs.~\ref{fig:o1} and \ref{fig:o1}, which are stable. 
Their energies relative to the symmetry unbroken state is given in Tab.~\ref{tab:t}. 
The remaining orders are not stable in this parameter regime. 
From these two orders, the stripe order type I is energetically favoured and therefore represents the true electronic ground state. 
A detailed mean field theory of this order including the critical interaction strength as a function of moiré band filling will be given in the next section. 

\begin{table}
	\begin{tabular}{r"c|c|c|c|c"c|c|c|c|c"c|c|c|c|c"c|c|c|c|c}
			$\epsilon-\epsilon_0$ in \negthinspace meV & \multicolumn{5}{c"}{$\beta=0.04$} & \multicolumn{5}{c"}{$\beta=0.06$} & \multicolumn{5}{c"}{$\beta=0.08$} & \multicolumn{5}{c}{$\beta=0.1$}\tabularnewline
		\hline 
		band filling $\nu$ & \negthinspace 0.6 & \negthinspace 0.55 & \negthinspace0.5 &\negthinspace 0.45 & \negthinspace0.4\negthinspace & 0.6 & 0.55 & 0.5 & 0.45 & 0.4 & 0.6 & 0.55 & 0.5 & 0.45 & 0.4 & 0.6 & 0.55 & 0.5 & 0.45 & 0.4\tabularnewline
		\thickhline 
		Stripe \negthinspace order \negthinspace I & - & - & - & - & - & -3.21 & -3.63 & -4.36 & -2.41 & -2.06 & -6.62 & -7.72 & -9.19 & -6.03 & -5.53 & -9.36 & -11.12 & -14.56 & -9.89 & -9.04\tabularnewline
		Stripe \negthinspace order \negthinspace II& - & - & - & - & - & -0.23 & -0.19 & - & 0.02 & -0.69 & -0.96 & -1.07 & -1.21 & -1.81 & -3.17 & -1.78 & -3.58 & -3.91 & -3.81 & -6.30\tabularnewline
	\end{tabular}
\caption{\label{tab:t}Energy per superlattice unit cell $ \epsilon $ of the stripe orders I and II depicted in Fig.~\ref{fig:order} relative to the energy of the symmetry unbroken state $ \epsilon_0 $ for various interaction strengths $ \beta $ and moiré band fillings $ \nu $.  }
\end{table}


\section{Mean field theory of charge-density wave order}
\label{app:mean field-theory}
In what follows, a detailed mean field theory of the charge density wave order
identified in Subsec.~\ref{sec:weak-coupling-regime}  as the true ground state of the weak-coupling regime is developed. 
The order breaks rotational and translational symmetry specified by the order parameter $ \Delta_\textbf{Q} $ with the ordering vector $ \textbf{Q} $ given in Eq.~\eqref{eq:orderPar},
but preserves spin and valley symmetry. 
Thus, only direct interaction channels including on-site, nearest neighbor- (NN),
next-to-nearest neighbor- (NNN) and next-to-next-to-nearest neighbor
(NNNN) interactions as discussed in Subsec.~\ref{subsec:interaction-ME} contribute to a formation of this particular order.
To determine its onset, i.e. the critical interaction strength for a given moiré band filling and temperature, two aspects have to be considered: The effective interaction strength in the charge density wave channel $ U_\text{CDW} $ and the corresponding static charge susceptibility $ \chi_\textbf{Q} $ with finite momentum transfer. Both enter the usual criterion for the onset of mean field orders (see e.g.~Ref.~\cite{Fazekas}), 
\begin{equation}
\label{eq:mf-criterion}
1+ [U_{\text{CDW}}\chi_{\mathbf{Q}}(\mu) ]\big|_\text{cr}=0,
\end{equation}
which determines the critical interaction strength. In the following, we will determine $ U_{\text{CDW}} $ and evaluate $ \chi_{\mathbf{Q}} $ to determine the critical effective interaction strength $ \beta_{\text{crit}} $.

\subsection{Mean field Hamiltonian}
We start with the interaction part of the Hamiltonian Eq.~\eqref{eq:eff-model} where only direct channels are considered,
\begin{equation}
\label{eq:intH-app}
H_{\text{int}}=\frac{1}{2}\sum_{ab}\sum_{\sigma\sigma'}U_{ab}c_{a\sigma}^{\dagger}c_{a\sigma}c_{b\sigma'}^{\dagger}c_{b\sigma'},
\end{equation}
where $a=(i,\alpha,\xi)$ contains the superlattice unit cell index $i$, the superlattice basis index
$\alpha\in\{AB,BA\}$, the valley index $\xi\in\{+,-\}$, and the spin
index $\sigma$. 
To determine the effective interaction strength,
Eq.~\eqref{eq:intH-app} is expressed in momentum space by using $c_{\mathbf{k}\alpha}=N^{-1/2}\sum_{i}e^{i\mathbf{R}_{i}\mathbf{k}}c_{i\alpha}$ with the lattice vector $ \mathbf{R}_i $, where
the valley and spin index is dropped for convenience but restored if necessary. 
By using $U_{ab}= \sum_l U_{l}\sum_{j}\delta_{\mathbf{R}_{a}-\mathbf{R}_{b},\mathbf{A}_{j}^{(l)}}$
where $\{\mathbf{A}_{j}^{(l)}\}$ denotes the set of space vectors
connecting the interacting lattice sites for the density interactions
of type $l\in\{\text{on-site},\text{NN},\text{NNN},\text{NNNN}\}$,
we obtain 
\begin{equation}
H_{\text{int}}^{(l)}=\frac{U_{l}}{2N}\sum_{\alpha\beta}\sum_{\mathbf{q}}\gamma_{\alpha\beta}^{(l)}(\mathbf{q})\rho_{\alpha}(\mathbf{q})\rho_{\beta}(-\mathbf{q}),
\end{equation}
where the density operator is given by $\rho_{\alpha}(\mathbf{q})=\sum_{\xi\sigma}\rho_{\alpha\xi\sigma}(\mathbf{q})$
with $\rho_{\alpha\xi\sigma}(\mathbf{q})=\sum_{\mathbf{k}}c_{\mathbf{k}+\mathbf{q}\alpha\xi\sigma}^{\dagger}c_{\mathbf{k}\alpha\xi\sigma}$
and the vertex function $\gamma_{\alpha\beta}^{(l)}(\mathbf{q})=\sum_{j}e^{i\mathbf{A}_{j}^{(l)}\mathbf{q}}$, which obeys  $[\gamma^{(l)}_{\alpha\beta}(\mathbf{q})]^{*}=\gamma^{(l)}_{\beta\alpha}(\mathbf{q})=\gamma^{(l)}_{\alpha\beta}(-\mathbf{q})$.
For non-local interactions, the vertex functions are determined to
\begin{subequations}
	\begin{align}
	\gamma_{AB}^{(\text{NN})}(\mathbf{q}) & =e^{-i \mathbf{u}_{AB}\mathbf{q}}(1+e^{i\mathbf{q}\mathbf{A}_{1}}+e^{i\mathbf{q}\mathbf{A}_{2}}),\\
	\gamma_{\alpha\alpha}^{(\text{NNN})}(\mathbf{q}) & =2\{\cos[\mathbf{q}\mathbf{A}_{1}]+\cos[\mathbf{q}\mathbf{A}_{2}]+\cos[\mathbf{q}(\mathbf{A}_{1}+\mathbf{A}_{2})]\},\\
	\gamma_{AB}^{(\text{NNNN})}(\mathbf{q}) & =e^{-i\mathbf{u}_{AB}\mathbf{q}}[e^{i\mathbf{q}(\mathbf{A}_{1}+\mathbf{A}_{2})}+e^{i\mathbf{q}(\mathbf{A}_{1}-\mathbf{A}_{2})}+e^{-i\mathbf{q}(\mathbf{A}_{1}-\mathbf{A}_{2})}],
	\end{align}
\end{subequations} 
with the Bravais lattice vector $ \textbf{A}_i $ and the basis vector $ \textbf{u}_{AB} $ which connects the crystalline basis sites. 
For transferred momenta $\mathbf{Q} \in \{\mathbf{G}_{1}/2,\mathbf{G}_{2}/2,\left(\mathbf{G}_{1}+\mathbf{G}_{2}\right)/2\} $,
the interaction part reduces to 
\begin{multline}
H_{\text{int}}=\tfrac{1}{N}U_{\text{on-site}}\Big[\sum_{\alpha\xi}\rho_{\alpha\uparrow\xi}(\mathbf{Q})\rho_{\alpha\downarrow\xi}(\mathbf{Q})+\sum_{\alpha\sigma}\rho_{\alpha\sigma+}(\mathbf{Q})\rho_{\alpha\sigma-}(\mathbf{Q})\Big]\\
+\tfrac{1}{2N}\left[U_{\text{NN}}-3U_{\text{NNNN}}\right]\rho_{A}(\mathbf{Q})\rho_{B}(\mathbf{Q})-U_{\text{NNN}}\sum_{\alpha}\rho_{\alpha}(\mathbf{Q})\rho_{\alpha}(\mathbf{Q}),
\end{multline}
where the on-site interaction entered with a constant vertex function. 
Due to the finite momentum transfer, we find negative interaction amplitudes 
for direct interactions of NNN- and NNNN-type.
This is traced back to the fact that for a developed charge density wave order
with ordering vector $\mathbf{Q}$ interaction contributions from these types of
interactions are minimized which can be inferred from the representation of a possible order depicted in Fig.~\ref{fig:weak-coupling-results}. 
Indeed, $\gamma_{\alpha\alpha}^{(\text{NNN})}(\mathbf{q})$
and $\gamma_{AB}^{(\text{NNNN})}(\mathbf{q})$ are minimal and negative
for $\mathbf{q}=\mathbf{Q}$. Thus, matrix elements $U_{\text{NNN}}$ and $U_{\text{NNNN}}$
favor the charge density wave order with ordering vector $\mathbf{Q}$, whereas $U_{\text{on-site}}$
and $U_{\text{NN}}$ act against it. 

The mean field Hamiltonian is obtained by introducing mean fields
$\langle\rho_{\alpha\sigma\xi}(\mathbf{Q})\rangle=\Delta_{\textbf{Q}}/8$ and dropping constant terms yielding 
\begin{align}
\label{eq:AppMFHamiltonian}
H_{\text{MF}} & =H_{0}+\tfrac{\Delta_{\textbf{Q}}}{4N}U_{\text{CDW}}\sum_{\alpha}\rho_{\alpha}(\mathbf{Q}),
\end{align}
where $H_{0}$ denotes the quadratic part of Eq.~\eqref{eq:eff-model}. The effective interaction strength  in the charge density wave channel is given by 
\begin{equation}
U_{\text{CDW}}=U_{\text{on-site}}+U_{\text{NN}}-4U_{\text{NNN}}-3U_{\text{NNNN}}.
\end{equation}

\begin{figure}[b]
	\begin{centering}
		\subfloat[\label{fig:charge-sus}]{\includegraphics[width=0.48\linewidth]{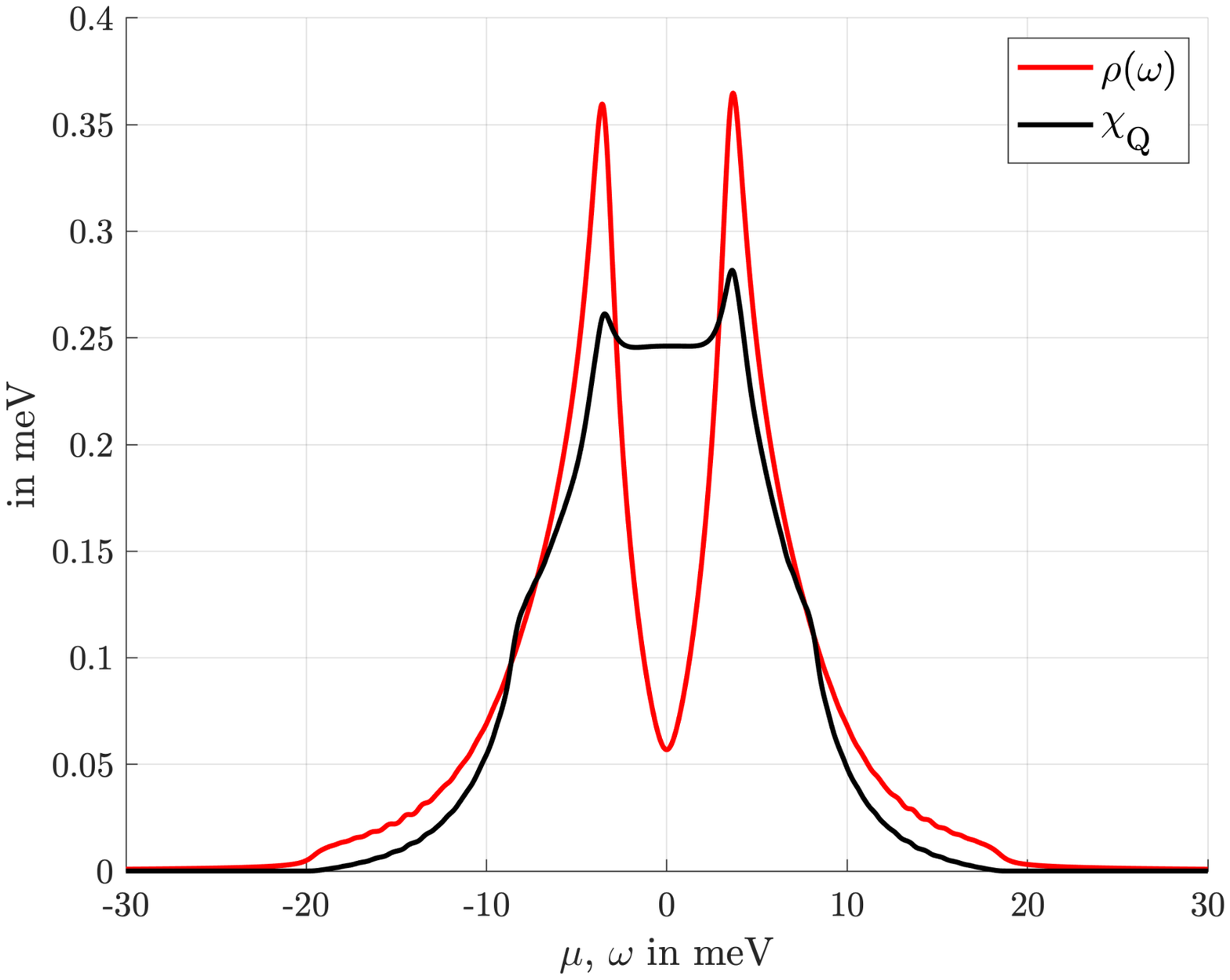}}
		\subfloat[\label{fig:betacrit}]{\includegraphics[width=0.48\linewidth]{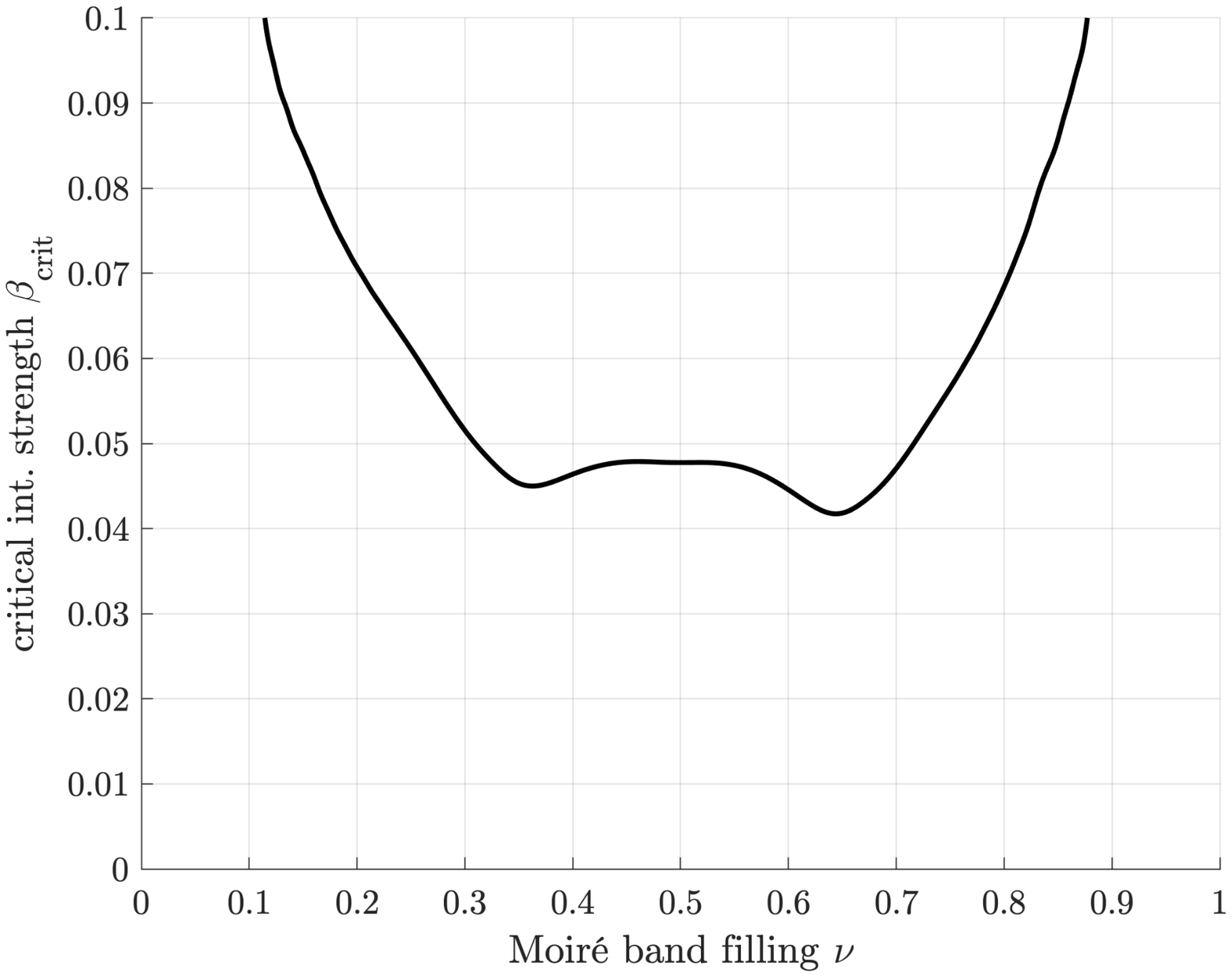}}
		\caption{\label{fig:CDWsus}(a) Charge susceptibility (black line) $ \chi_\textbf{Q}(\mu) $ and single-particle density of states (red line) $ \rho(\omega) = -\frac{1}{\pi N} \sum_{\textbf{k}} \text{Im} G^R_0 (\omega,\textbf{k})\big|_{\mu=0} $ for a fixed temperature $k_B  T\approx0.1\Lambda $ as a function of the doping level and the frequency, respectively, for a twist-angle representative for the weak-coupling regime.
			(b) Critical interaction strength for a fixed temperature as a function of the moiré band filling. 
		}
		\par\end{centering}
\end{figure}

\subsection{Critical interaction strength}
Besides the effective interaction strength $ U_\text{CDW} $, the criterion Eq.~\eqref{eq:mf-criterion}, which determines the onset of the density wave order, is determined by the respective 
charge susceptibility $ \chi_{\mathbf{Q}} $ with finite momentum and zero frequency transfer. 
By means of standard methods (see e.g.~Ref.~\cite{Fazekas}), it is determined to 
\begin{equation}
\chi_{\mathbf{Q}}(\mu)=-\frac{i}{4N}\sum_{\mathbf{k}}\int_{}^{} \frac{d\omega}{2\pi} 
f(\omega-\mu)\text{tr}_{\alpha\xi\sigma}\big[VG_{0,\mathbf{k}}^{R}(\omega)VG_{0,\mathbf{k}+\mathbf{Q}}^{R}(\omega)-VG_{0,\mathbf{k}}^{A}(\omega)VG_{0,\mathbf{k}+\mathbf{Q}}^{A}(\omega)\big],
\end{equation}
where the free electronic Green's function is given by 
\begin{equation}
G_{0,\alpha\beta}^{R}(\omega,\mathbf{k})= \big([\omega-h_{0}(\mathbf{k})+i0^{+}]^{-1}\big)_{\alpha\beta},
\end{equation}
with $ h_0 $ obtained from the quadratic part of Eq.~(\ref{eq:eff-model}) by Fourier transform being diagonal in spin and valley space, the vertex part
$V_{\alpha\beta}=\delta_{\alpha\beta}$ and the Fermi function $  f(\omega) = [e^{\omega/k_B T}+1]^{-1}$. 
$ \chi_\textbf{Q} $ is evaluated numerically and is determined as function of the doping level $ \mu $ as depicted in Fig.~\ref{fig:charge-sus}. 

The charge susceptibility $ \chi_{\mathbf{Q}} $ is slightly peaked at the positions of the van Hove points but does not diverge for any doping level because of the absence of a nesting condition connected with the momentum transfer $ \textbf{Q} $. Although the density of states vanishes at the CNP, substantial weight of the susceptibility is also observed at the CNP because of the finite momentum transfer. 
By revisiting the condition for the onset of the mean field order Eq.~\eqref{eq:mf-criterion}, we deduce that there is a finite, attractive critical interaction strength for a given temperature and chemical potential (or moiré band filling). 
This finding is made more explicit by rearranging the criterion Eq.~\eqref{eq:mf-criterion} to determine the critical interaction strength at which the order develops.
$ \beta_{\text{crit}}  $ is plotted as function of moiré band filling $ \nu $ in Fig.~\ref{fig:betacrit}.  
Additionally, we determine the order parameter $ \Delta_\textbf{Q} $ defined in Eq.~\eqref{eq:orderPar} for a fixed temperature as a function of $ \beta $
by solving the gap equation Eq.~\eqref{eq:corrs}, but now for the mean field Hamiltonian of the stripe charge density wave given in Eq.~\eqref{eq:AppMFHamiltonian}. 
The numerical evaluation of the gap equation is conducted in reciprocal space, which is much more efficient and allows for higher resolved results.  
The results are depicted in Fig.~\ref{fig:orderParameter}.
We observe that 
the amplitude of the order parameter deep in the symmetry broken phase is largest for half-filling, whereas $ \beta_{\text{crit}} $ is smallest for the moiré band filling $ \nu \approx 0.63 $ in the vicinity of the van Hove peaks of the single-particle spectrum which is inline with the results for $ \beta_{\text{crit}} $ shown in Fig.~\ref{fig:betacrit}. 
\begin{figure}[b]
	\begin{centering}
		\includegraphics[width=0.55\linewidth]{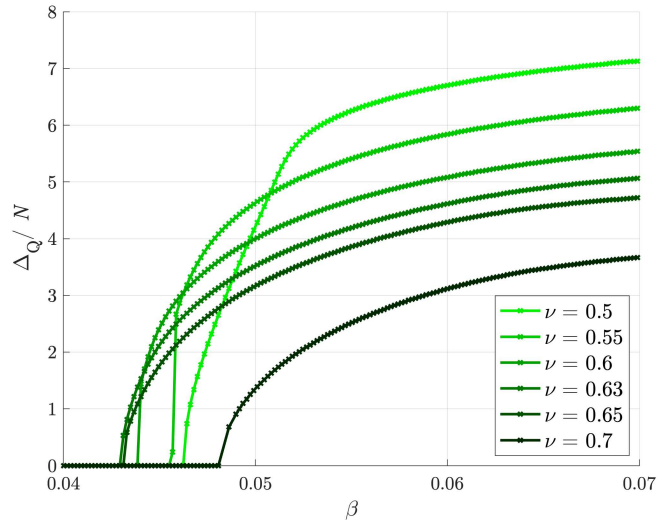}
		\caption{\label{fig:orderParameter}Order parameter $ \Delta _\mathbf{Q}$ of the striped charge density wave for a fixed temperature $ k_B T\approx0.1\Lambda $ as a function of the effective coupling strength $ \beta $. 
		}
		\par\end{centering}
\end{figure}

Hence, for a finite attractive interaction in the charge density wave channel $ U_\text{CDW} < 0 $ whose critical value is determined by Eq.~\eqref{eq:mf-criterion}, an onset of the charge density wave order is expected. 
The amplitude of the order parameter is expected to be largest for doping levels around charge neutrality with lowest critical interaction strengths near the von Hove points of the single-particle spectrum as depicted in Fig.~\ref{fig:results}. 

\section{Strong coupling analysis \label{app:sc}}
In the strong coupling regime, we consider the Hamiltonian $ H_\text{SC} $ specified in Eq.~\eqref{eq:HSC}. As $ [H_\SC,\hat{n}_{a\sigma}] = 0 $, the local densities represent conserved quantities rendering the model a classical model which can be solved by employing classical methods. Hence, the partition function is given by 
\begin{equation}
Z = \prod_{i=1}^N \Big(\sum_{n_i=0}^1\Big)  e^{-\frac{E_\SC[n]}{k_B T}} 
\end{equation}
where $ N $ specifies the total number of sites which are labelled by $ i=\{a,\sigma\} $ and $ n_i \in \{0,1\} $ the local occupation number. The energy associated with the state $ n = \{n_1,\dots,n_N\} $ is given by 
\begin{equation}
	E_\SC[n] = \tfrac{1}{2}\sum_{\sigma\sigma'}\sum_{ab\in\hexagon}(U_{ab}-J_{ab}\delta_{\sigma\sigma'}) (n_{a\sigma}-\tfrac{1}{2})(n_{b\sigma'}-\tfrac{1}{2}). 
\end{equation}
Accordingly, thermal expectation values are given by $ \langle \hat{O} \rangle =  Z^{-1} \prod_{i=1}^N \big(\sum_{n_i=0}^1\big) O_n  e^{-\frac{E_\SC[n]}{k_B T} }  $, where $ O_n $ is the value of the observable for a particular state specified by the configuration $ n $. 

In what follows, we are interested in the ground state configuration $ n_\text{GS} $ which minimizes the energy functional $ E_\SC[n] $, i.e.~
\begin{equation}
 E_\SC[n_\text{GS} ] \equiv \min_n E_\SC[n],  
\end{equation}
such that $ \langle H_{\text{SC}} \rangle|_{T\rightarrow 0} = E_\SC[n_\text{GS}]  $.
To determine $ n_\text{GS} $, we employ the simulated annealing algorithm \cite{Kirkpatrick83:SimulatedAnnealing}. 
It is a Monte Carlo-based optimization algorithm which is suited for high-dimensional problems and which effectively scans the available state space. This method is standard and can be found, e.g., in Ref.~\cite{Laarhoven87:SimAnnealingBook}. Our results for $ n_\text{GS} $ for various commensurate moiré band fillings are depicted in Fig.~\ref{fig:sc-results}. 

\end{widetext}
	
\end{document}